\documentclass[apj,numberedappendix,appendixfloats]{emulateapj}
\usepackage{apjfonts}

\newcommand{\etal}{et~al.}
\newcommand{\MgIIdblt}{{\rm Mg}\kern 0.1em{\sc ii}~$\lambda\lambda 2796, 2803$}
\newcommand{\MgII}{\hbox{{\rm Mg}\kern 0.1em{\sc ii}}}
\newcommand{\HI}{\hbox{{\rm H}\kern 0.1em{\sc i}}}
\newcommand{\CIV}{\hbox{{\rm C}\kern 0.1em{\sc iv}}}
\newcommand{\OVI}{\hbox{{\rm O}\kern 0.1em{\sc vi}}}
\newcommand{\Lya}{\hbox{{\rm Ly}\kern 0.1em $\alpha$}}
\newcommand{\kms}{\hbox{km~s$^{-1}$}}

\newcommand{\gcc}{\hbox{g~cm$^{-3}$}}

\newcommand{\magiicat}{\hbox{{\rm MAG}{\sc ii}CAT}}

\slugcomment{Accepted for publication in ApJ, September 11, 2013}
\shorttitle{\sc ~{\magiicat}}
\shortauthors{\sc Nielsen {\etal}}

\begin{document}

\title{~{\magiicat} I. The {\MgII} Absorber-Galaxy Catalog}

\author{\sc
Nikole M. Nielsen\altaffilmark{1},
Christopher W. Churchill\altaffilmark{1},
Glenn G. Kacprzak\altaffilmark{2,3},
and
Michael T. Murphy\altaffilmark{2}
}

\altaffiltext{1}{New Mexico State University, Las Cruces, NM 88003
{\tt nnielsen@nmsu.edu}}
                                                                                
\altaffiltext{2}{Swinburne University of Technology, Victoria 3122,
Australia}

\altaffiltext{3}{Australian Research Council Super Science Fellow}

\begin{abstract}

We describe the {\MgII} Absorber-Galaxy Catalog, {\magiicat}, a
compilation of 182 spectroscopically identified intermediate redshift
($0.07\leq z\leq 1.1$) galaxies with measurements of {\MgIIdblt}
absorption from their circumgalactic medium within projected distances
of 200~kpc from background quasars. In this work, we present
``isolated'' galaxies, which are defined as having no
spectroscopically identified galaxy within a projected distance of
$100$~kpc and a line of sight velocity separation of $500$~{\kms}. We
standardized all galaxy properties to the $\Lambda$CDM cosmology and
galaxy luminosities, absolute magnitudes, and rest-frame colors to the
$B$- and $K$-band on the AB system. We present galaxy properties and
rest-frame {\MgII} equivalent width, $W_r(2796)$, versus galaxy
redshift. The well-known anti-correlation between $W_r(2796)$ and
quasar-galaxy impact parameter, $D$, is significant to the $8~\sigma$
level. The mean color of {\magiicat} galaxies is consistent with an
Sbc galaxy for all redshifts. We also present $B$- and $K$-band
luminosity functions for different $W_r(2796)$ and redshift
subsamples: ``weak absorbing'' [$W_r(2796)<0.3$~{\AA}], ``strong
absorbing'' [$W_r(2796)\geq 0.3$~{\AA}], low redshift ($z<\langle
z\rangle$), and high redshift ($z\geq \langle z\rangle$), where
$\langle z\rangle =0.359$ is the median galaxy redshift. Rest-frame
color $B-K$ correlates with $M_K$ at the $8~\sigma$ level for the
whole sample but is driven by the strong absorbing, high redshift
subsample ($6~\sigma$). Using $M_K$ as a proxy for stellar mass and
examining the luminosity functions, we infer that in lower stellar
mass galaxies, {\MgII} absorption is preferentially detected in blue
galaxies and the absorption is more likely to be weak.

\end{abstract}

\keywords{galaxies: halos --- quasars: absorption lines}

\section{Introduction}
\label{sec:intro}

Galaxies are known to harbor large, extended reservoirs of gas
referred to as the circumgalactic medium (CGM). This region contains
the material through which filaments accrete, galactic-scale winds
outflow, and merging galaxies are tidally stripped -- mechanisms which
are critical to the growth and transformation of galaxies given that
the CGM harbors a gas mass which may rival that of the galaxy itself
\citep{tumlinson11,stocke13,werk13}. Additionally, theoretical works
have established that the baryons in the CGM depend on the dark matter
halo mass and various processes such as stellar and active galactic
nucleus feedback \citep{birnboim03, maller04, keres05, dekel06,
  birnboim07, ocvirk08, keres09, oppenheimer10, stewart11,
  vandevoort11, vandevoort+schaye11}. As such, the evolution of
galaxies is intimately tied to the origin, processing, and fate of gas
in their halos, making studies of the CGM key to understanding galaxy
evolution.

Campaigns to study the CGM in absorption at $z \leq 1$ have targeted
various ions that probe a range of gas densities and temperatures. For
$z<0.3$, {\OVI} absorption \citep[e.g.,][]{tumlinson11,stocke13}
traces gas with $n_{\hbox{\tiny H}} \sim 10^{-4}$~{\gcc} between
$T=10^{4.8}$~K (photoionized) and $T=10^{5.5}$~K (collisionally
ionized). At $z<1$, {\CIV} absorption \citep[e.g.,][]{chen01a} probes
$10^{-2} \leq n_{\hbox{\tiny H}} \leq 10^{-4}$~{\gcc} gas with
temperatures in the range of $\sim 10^{4.6}$~K (photoionized) and
$\sim 10^{5.0}$~K (collisionally ionized). The neutral hydrogen
component of the CGM has been observed using {\Lya} absorption
\citep[e.g.,][]{lanzetta95,chen01b,stocke13}. However, by far the vast
majority of surveys have focused on {\MgII} absorption
\citep[e.g.,][]{bb91, sdp94, cwc-china, chen10a, kcems}, which samples
photoionized CGM gas with $n_{\hbox{\tiny H}} \sim 10^{-1}$~{\gcc} and
$T \sim 10^{4.5}$~K. Further details of $z<1$ absorbing gas properties
are discussed in \citet{bergeron94}.

The {\MgIIdblt} absorption doublet is well-suited to studying the
processes occurring in the CGM since it is easily observed from the
ground in the optical at redshifts $0.1<z<2.5$. {\MgII} traces
metal-enriched, low ionization gas over a large range of {\HI} column
densities, $16\lesssim \log N({\HI})\lesssim 22$ \citep{bs86, ss92,
  weakI, archiveI, rao00, weakII}, corresponding to a wide range of
environments out to projected distances of $\sim 150$~kpc
\citep{ggk08, chen10a, churchill-weakgals}. Detailed information on
the gas kinematics with {\MgII} have indicated the presence of
infalling gas \citep{ggk-sims, kcems, ribaudo11, stewart11, martin12,
  rubin12, ggk-1317} and outflowing galactic-scale winds
\citep{bouche06, tremonti07, martin09, weiner09, chelouche10, rubin10,
  bordoloi11, coil11, bouche12, martin12}.

The various methods employed for surveys of {\MgII} absorbing galaxies
present challenges in understanding the CGM-galaxy interaction. The
largest survey has no more than $\sim 80$ isolated galaxies, yet some
200 are known. Galaxy absolute photometric properties and
quasar-galaxy impact parameters were computed using a variety of
cosmological parameters (the accepted cosmology at the time a given
survey was published), which have changed over the last $\sim 20$
years. Different observing facilities have been used, resulting in
various filter sets. Different magnitude systems were employed. Even
the selection methods from survey to survey are diverse, and in some
cases different surveys report the same absorber-galaxy pairs, causing
duplicates throughout the literature. All these factors result in
difficulties when synthesizing the galaxy properties and results
between studies.

Motivated by the potential that combining the data from our work and
other surveys may further illuminate the CGM-galaxy connection, we
have endeavored to assemble a database of the extant works focused on
{\MgII} absorption from the CGM of intermediate redshift galaxies. We
aim to provide a large uniform data suite based upon a single
cosmological parameter set and to standardize all absolute magnitudes
to two filters on the AB system. We consolidate the measurements of
given absorber-galaxy pairs duplicated in various works to include the
highest-quality data available for each absorber-galaxy pair. Such a
compilation holds the promise of yielding higher statistical
significance in the already published results, and of providing
greater leverage for exploring the dependence of CGM {\MgII}
absorption on various galaxy properties.

In this paper we present the construction of the {\MgII}
Absorber-Galaxy Catalog, {\magiicat} (pronounced magic-cat). We also
present the data and general characteristics of the ``isolated
galaxy'' sample. In several works from which the absorber-galaxy
pairs were drawn, group galaxies (which we define below) were
identified. Since the absorption associated with multi-galaxy pairs
may be probing the intragroup medium, or providing a single absorption
measurement from the overlap of the circumgalactic medium from more
than one galaxy, we defer presentation and analysis of these pairs for
future work. Our aim here is to focus on the CGM-galaxy connection
for cases in which the data are consistent, to the best of our
knowledge, with the absorption arising in the CGM of a single dominant
host galaxy. In a future paper, we will present the ``group galaxy''
subsample of {\magiicat} and will examine the intragroup medium
environment.

We also leave further detailed analysis of the CGM-galaxy connection
for other papers in this series, e.g., Paper~II \citep{nielsen12}, in
which we studied the general characteristics of the CGM with galaxy
luminosity, color, and redshift, and Paper~III \citep{cwc-masses2}, in
which we studied the behavior of the CGM with galaxy virial mass. An
additional series paper is planned in which we will study the
kinematics of the {\MgII} absorbing CGM. We have also presented
additional analysis of the ``isolated'' galaxy subsample in
\citet{kcn12} and \citet{churchill-masses}.

In \S~\ref{sec:construct} we provide the selection criteria for
inclusion of galaxies in the catalog and briefly describe each of the
works from which the galaxies are drawn and the various selection
methods. In \S~\ref{sec:methods} we detail the galaxy data we obtained
and how we standardized various galaxy and absorption properties. We
adopt a $\Lambda$CDM cosmology ($H_0=70$ km s$^{-1}$ Mpc$^{-1}$,
$\Omega_M=0.3$, and $\Omega_{\Lambda}=0.7$) and report AB absolute
magnitudes throughout this paper. In \S~\ref{sec:sample} we present
characteristics of the sample, tabulated values for {\magiicat}
galaxies, and luminosity functions. We summarize the present work and
conclude with the potential of {\magiicat} in
\S~\ref{sec:conclusions}. The catalog is available in its entirety in
the on-line journal and has been placed on-line at the NMSU Quasar
Absorption Line Group
website\footnote{http://astronomy.nmsu.edu/cwc/Group/magiicat}.

\section{Constructing {\magiicat}}
\label{sec:construct}

We compiled a catalog of galaxies with spectroscopic redshifts $0.07
\leq z \leq 1.1$ within a projected distance of $D \leq 200$~kpc from
a background quasar, with known {\MgII} absorption or an upper limit
on absorption less than $0.3$~{\AA}. We chose to include only galaxies
with spectroscopic redshifts, excluding galaxies from e.g.,
\citet{rao11} and \citet{bowen11}, which have photometric redshifts
and would have supplied $\sim 30$ and $\sim 10$ galaxies to
{\magiicat}, respectively. We also limited the sample to galaxies
which are not located in group environments (defined in
\S~\ref{sec:groupgals}) to the limits the data indicate. The galaxies
were primarily drawn from the works of \citet{sdp94}, \citet{csv96},
\citet{guillemin}, \citet{steidel97}, \citet{chen08},
\citet{barton09}, \citet{chen10a}, \citet{ggk1127},
\citet{gauthier11}, \citet{kacprzak11kin,kcems}, and
\citet{churchill-weakgals}.

The galaxy discovery methods employed by the aforementioned surveys
range from unbiased volume-limited samples with no {\it a priori\/}
knowledge of {\MgII} absorption in the background quasar spectrum
\citep{barton09,gauthier11,kacprzak11kin}, to magnitude-limited
samples \citep{steidel97,ggk1127}, one with a luminosity scaled
maximum projected separation from the quasar sightline
\citep{chen10a}, and to samples in which galaxies are searched for at
the redshifts of known {\MgII} absorbers \citep[i.e., absorption
  selected;][]{bb91,sdp94,guillemin,chen08,gauthier11,kcems}. Some
quasar fields have been imaged from the ground only (some with and
some without subtraction of the quasar), while others have been imaged
at high resolution with the {\it Hubble Space Telescope} ({\it
  HST}). As such, our compilation comprises a catalog of galaxies with
heterogeneous selection methods, and a range of sensitivity in
magnitude and impact parameter. Though it may be argued that a
complete galaxy sample is indicated by always identifying a galaxy at
the redshift of known {\MgII} absorption \citep{sdp94,steidel95}, it
is inherently difficult to demonstrate completeness unless the quasar
fields are systematically surveyed to a uniform magnitude limit and
projected separation from the quasar.

\subsection{Overview of Surveys}
\label{sec:overview}

Here we present a brief overview of the previous works included in
{\magiicat}. 

\subsubsection{SDP94} 

We obtained the data for galaxies presented in \citet{sdp94}
[hereafter SDP94] with $0.3\leq z \leq 1.0$ (Steidel, private
communication). Their sample is ``gas cross section-selected,''
meaning that the galaxies were selected based on known {\MgII}
absorption with rest-frame equivalent widths $W_r\geq 0.3$~{\AA} in
the spectra of background quasars. Galaxies were searched for starting
at the quasar position and moving outward in angular separation,
$\theta$, with most galaxies having an angular separation less than
$10''$. Images of the quasar fields were acquired in the
$R_s$ band\footnote{Chuck Steidel kindly provided the electronic versions
 of the $R_s$ and $K_s$ filter response curves.} using the 2.1~m
and 4~m telescopes at Kitt Peak National Observatory, as well as the
2.4~m Hiltner telescope at the Michigan-Dartmouth-MIT
Observatory. Images in the infrared $K_s$ band were obtained with
NICMOS III cameras on the Kitt Peak 4~m Mayall telescope and the Las
Campanas Observatory 2.5 m DuPont telescope. Galaxy spectroscopy was
conducted using the Lens/Grism Spectrograph and the Kast Double
Spectrograph on the Lick Observatory 3~m Shane telescope. Roughly 30\%
of the galaxies identified by SDP94 do not have spectroscopically
confirmed redshifts; we did not include those galaxies. Many of the
galaxies from SDP94 were studied more extensively in later works, and
are therefore listed under the most recent work. 

\subsubsection{Steidel-PC} 

Steidel (private communication) kindly provided the unpublished
``interloper'' galaxy data briefly discussed as ``control fields'' in
SDP94 and \citet{steidel95}. These galaxies were targeted because they
were not responsible for absorption [to a $5~\sigma$ upper limit
 $W_r(2796)<0.3$~{\AA}] in background quasar spectra during the
campaign of SDP94. Galaxy images and spectroscopy were obtained using
the same facilities used by SDP94. Later works have obtained
HIRES/Keck \citep{vogt94} or UVES/VLT \citep{dekker00} quasar spectra
for all galaxies in this sample, therefore the equivalent width limits
for these galaxies have been remeasured at the $3~\sigma$ level by
\citet{churchill-weakgals} or the present work. 

\subsubsection{GB97} 

Studying quasar fields with known {\MgII} absorbers, \citet{guillemin}
[hereafter GB97] identified galaxies producing the absorption at
$0.07<z<1.2$, with $R<23.5$ and quasar-galaxy angular separations
$\theta < 15''$. Imaging in the $R$ band\footnote{The electronic
  version of the ESO Faint Object Spectrograph and Camera $R$ filter
  response curve was kindly provided by Jacqueline Bergeron.} and
galaxy spectroscopy were conducted on the European Southern
Observatory 3.5~m telescope using the ESO Faint Object Spectrograph
and Camera (hereafter the $R$-band from this work will be referred to
as $R_{\hbox{\tiny EFOSC}}$). We did not include the high redshift
{\it candidate} absorbers from GB97.

\subsubsection{Steidel97} 

\citet{steidel97} [hereafter Steidel97] conducted a deep,
magnitude-limited study of the overdense galaxy field within $\theta =
50''$ of 3C 336 (1622+238). Galaxies as faint as $R_s=24.5$ were
imaged in $R_s$ with the Michigan-Dartmouth-MIT 2.4~m Hiltner
telescope and F702W with WFPC2 on the {\it Hubble Space Telescope}
({\it HST}). Infrared $K_s$ band images were obtained with the Kitt
Peak 4~m Mayall telescope, with the final image reaching Vega
magnitude $K_s\sim 22$ (AB magnitude $K_s\sim 23.8$). Quasar spectra
were collected from various instruments and telescopes: the Faint
Object Spectrograph on {\it HST}, the Kast Double Spectrograph with
the Lick Observatory 3~m Shane telescope, the RC Spectrograph at the
Kitt Peak 4~m Mayall telescope, and the Low Resolution Imaging
Spectrograph (LRIS) on the 10 m Keck-I telescope. Galaxy spectroscopy
was conducted with the LRIS/Keck-I combination. The data presented in
Steidel97 have been improved upon in later works, in fact a UVES/VLT
spectrum is now available, therefore we reference these galaxies according
to the work giving updated values, with Steidel97 as the source
of the $K_s$ magnitudes. 

\subsubsection{CT08} 

\citet{chen08} [hereafter CT08] selected several quasar fields for
which about half of the galaxy data was available from prior {\MgII}
surveys (Steidel97), and half from {\Lya} and {\CIV} surveys
\citep{lanzetta95,chen98,chen01a,chen01b}. The majority of galaxies
were imaged in the F702W band with WFPC2 on {\it HST}, while galaxies
in the field 0226-4110 were imaged in $R_J$ with IMACS on the Magellan
Baade telescope. Quasar spectroscopy was obtained with the MIKE
Echelle Spectrograph on the Magellan Clay telescope or were obtained
from the ESO data archive where they had been observed with UVES on
the Very Large Telescope (VLT).

\subsubsection{BC09} 

Working with the Sloan Digital Sky Survey (SDSS) Data Release 4 (DR4)
and Data Release 6 (DR6), \citet{barton09} [hereafter BC09] performed
a volume-limited survey of galaxies at $z\sim0.1$, with a limiting
absolute magnitude $M_r\leq -21.3$. Background quasars were selected
at projected distances less than or equal to 107~kpc from the galaxies
from the SDSS Data Release 6 (DR6) quasar catalog. Galaxy spectra were
collected from SDSS while quasar spectra were obtained using the blue
channel of LRIS on Keck-I. We obtained apparent SDSS ``model'' $g$ and
$r$ magnitudes from a NASA/IPAC Extragalactic Database
(NED)\footnote{http://ned.ipac.caltech.edu} search. The equivalent
widths for several of these galaxies were remeasured by
\citet{kacprzak11kin}.

\subsubsection{Chen10} 

\citet{chen10a} [hereafter Chen10] photometrically selected galaxies
with $z<0.5$ from the SDSS DR6 archive with $r'<22$. Galaxies in
quasar fields were targeted with the limitation that the quasar-galaxy
impact parameter, $D$, must be less than the expected gaseous radius
for each galaxy, $R=R_{\ast}(L_B/L_B^{\ast})^{0.35}$, where
$R_{\ast}=130$~kpc. Follow up galaxy and quasar spectroscopy was
obtained using the Dual Imaging Spectrograph (DIS) on the 3.5~m
telescope at the Apache Point Observatory (APO) or with MagE on the
Magellan Clay Telescope at the Las Campanas Observatory. We obtained
apparent SDSS model $g$ and $r$ magnitudes from NED. The published
$W_r(2796)$ upper limits were converted from $2~\sigma$ to $3~\sigma$
upper limits and we included only those galaxies for which the upper
limit was less than or equal to 0.3~{\AA}.

\subsubsection{KMC10} 

Performing a magnitude-limited survey down to F814W$\leq 20.3$,
\citet{ggk1127} [hereafter KMC10] studied the field 1127-145 which
contains many bright galaxies within an angular quasar-galaxy
separation of $50''$. The field was imaged in the F814W band on {\it
  HST} with WFPC2. Spectroscopy of the galaxies was obtained with DIS
at the APO 3.5~m telescope, while quasar spectroscopy was conducted
with UVES/VLT.

\subsubsection{GC11} 

\citet{gauthier11} [hereafter GC11] studied luminous red galaxies
(LRGs) that were photometrically identified in SDSS DR4 with $z\approx
0.5$ and a maximum $D$ corresponding to the fiducial virial radius of
LRGs, which is given as 500 kpc. A subset of their LRGs were selected
by known {\MgII} absorbers in quasar spectra, while a larger subset
was randomly selected near quasar sightlines with no prior knowledge
of {\MgII} absorption. Galaxy spectra were obtained using DIS on the
3.5~m telescope at APO or with the Boller \& Chivens Spectrograph on
the Las Campanas DuPont telescope. We obtained apparent SDSS ``model''
$g$ and $r$ magnitudes from NED and included only those LRGs with
impact parameters $D<200$~kpc in {\magiicat}. We also converted the
published upper limits on $W_r(2796)$ from $2~\sigma$ to $3~\sigma$,
including only those galaxies with $W_{\rm lim} \leq 0.3$~{\AA}.

\subsubsection{KCBC11} 

\citet{kacprzak11kin} [hereafter KCBC11] built upon the work of BC09
by adding galaxies of somewhat larger redshift, $z\sim 0.13$. Galaxy
spectra were obtained with DIS on the 3.5~m APO telescope, while
quasar spectra were collected using the blue channel of LRIS on
Keck-I. Using NED, we obtained the SDSS ``model'' $r$ and 2MASS total
$K_s$ apparent magnitudes for all but two galaxies, where $K_s$ was
not available. We instead obtained the SDSS ``model'' $g$ and $r$
magnitudes for those two galaxies.

\subsubsection{KCEMS11} 

Galaxies selected by \citet{kcems} [hereafter KCEMS11] are based on
known {\MgII} absorption at $0.3<z<1$ in quasar spectra, with many
having been originally identified by SDP94. Imaging of the quasar
fields was conducted by {\it HST} with the F702W or F814W filters on
WFPC2. Infrared imaging for the SDP94 fields was conducted in the
$K_s$ band with the NICMOS III cameras on the Kitt Peak 4~m Mayall
telescope and the Las Campanas Observatory 2.5~m DuPont
telescope. $K'$ magnitudes were obtained for three galaxies from
\citet{chen01b}, who used NSFCAM on the IRTF 3 m telescope, and F160W
magnitudes for two galaxies from David Law (private communication) who
observed them with WFC3 on {\it HST}. Quasar spectroscopy was obtained
with HIRES/Keck or UVES/VLT, while galaxy spectroscopic redshifts were
collected from the literature (see Table 3 in KCEMS11).

\subsubsection{Churchill13} 

Requiring high resolution quasar spectra ($R=45,000$) and {\it HST}
images, \citet{churchill-weakgals} [hereafter Churchill13] studied
galaxies with weak [$W_r(2796)<0.3$~{\AA}] or undetected {\MgII}
absorption. Half of the galaxies in this sample were obtained from
Steidel-PC, while the other half were collected from Steidel97,
\citet{cwc1317}, CT08, and KMC10. Galaxies were imaged in F702W or
F814W with WFPC2 on {\it HST} and, in some cases, in $K_s$ as detailed
in Steidel-PC. Quasar spectra were obtained with HIRES/Keck, UVES/VLT,
or with the MIKE Echelle Spectrograph on the Magellan Clay
telescope. We included galaxies listed as the ``New Sample'' in Table
2 in Churchill13.

\subsection{Sources of Data}

In Table~\ref{tab:surveys}, we present the sources of data for
galaxies in {\magiicat}. We present the work from which galaxies were
obtained in column (1). The rest of the columns contain the number of
galaxies for which the following data was obtained: (2) the galaxy
identification, such as redshift and impact parameter, (3) the
observed magnitude used to calculate the $B$-band absolute magnitude,
(4) the observed magnitude used to calculate the $K$-band absolute
magnitude, and (5) the {\MgII} equivalent width, while columns (6) and
(7) contain the number of galaxies that have detectable absorption and
have an upper limit on absorption, respectively.

We note that for many of the sources from which data were obtained,
the images of the quasar fields were not published. In cases in which
images were published and/or annotated, we did not have access to the
electronic data in order to verify and/or directly characterize the
magnitude limits and galaxy detection thresholds. As such, we have not
been able to directly inspect the imaged fields around the quasars for
a substantial portion of the sample. In addition, in many cases the
spectra showing the {\MgII} absorption lines were not published so
that we have had to rely upon the published equivalent width
measurements.

\begin{deluxetable}{lcccccc}
\tabletypesize{\footnotesize}
\tablecolumns{7}
\tablewidth{0pt}
\setlength{\tabcolsep}{0.06in}
\tablecaption{Sources of Data in {\magiicat}~\tablenotemark{a} \label{tab:surveys}}
\tablehead{
  \multicolumn{2}{c}{}                &
  \multicolumn{2}{c}{Magnitudes}      &
  \multicolumn{3}{c}{------------ Absorption ------------} \\
  \colhead{(1)}                       &
  \colhead{(2)}                       &
  \colhead{(3)}                       &
  \colhead{(4)}                       &
  \colhead{(5)}                       &
  \colhead{(6)}                       &
  \colhead{(7)}                       \\
  \colhead{Survey}                    &
  \colhead{Galaxy}                    &
  \colhead{$B$}                       &
  \colhead{$K$}                       &
  \colhead{$W_r(2796)$}               &
  \colhead{Absorbers}                 &
  \colhead{Upper}                     \\
  \colhead{}                          &
  \colhead{ID}                        &
  \multicolumn{4}{c}{}                &
  \colhead{Limits}                     }
\startdata
GB97        &   9 &   9 &   0 &   3 &   3 &   0 \\[3pt]
Churchill13 &  18 &  17 &   9 &  20 &   0 &  20 \\[3pt]
KCEMS11     &  33 &  33 &   0 &  32 &  32 &   0 \\[3pt]
Steidel97   &   0 &   0 &  11 &   0 &   0 &   0 \\[3pt]
CT08        &   5 &   5 &   0 &   3 &   2 &   1 \\[3pt]
Chen10      &  68 &   0 &   0 &  68 &  47 &  21 \\[3pt]
SDP94       &  18 &  18 &  38 &   7 &   7 &   0 \\[3pt]
Steidel-PC  &   0 &   1 &   1 &   0 &   0 &   0 \\[3pt]
KMC10       &   4 &   4 &   0 &   1 &   0 &   1 \\[3pt]
KCBC11      &   9 &   0 &   0 &   9 &   9 &   0 \\[3pt]
GC11        &  12 &   0 &   0 &  12 &   5 &   7 \\[3pt]
BC09        &   6 &   0 &   0 &   6 &   0 &   6 \\[3pt]
This Work   &   0 &  95 & 105 &  21 &  18 &   3 \\[3pt]
Total       & 182 & 182 & 164 & 182 & 123 &  59 \\[-5pt]
\enddata
\tablenotetext{a}{The numbers associated with each ``survey'' reflect
  the source from which the data we published are taken.  In cases
  where a ``0'' appears, this is because all of the galaxy or
  absorption data was either published elsewhere for the first time or 
  republished with higher quality data in a later work.}
\end{deluxetable}

\subsection{Defining Isolated and Group Galaxies}
\label{sec:groupgals}

In this paper, we present only ``isolated'' galaxies since we are
focused on the CGM of individual systems. We define an isolated
galaxy to be one in which there is no spectroscopically identified
galaxy within 100~kpc (projected) and a line of sight velocity
separation of 500~{\kms}. Conversely, a group galaxy is defined to
have a spectroscopically identified nearest neighbor within a
projected separation of 100~kpc and having a line of sight relative
velocity less than 500~{\kms}. This definition is adopted from
\citet{chen10a}, but modified to include a slightly larger velocity
separation.

Since we are unable to examine all images of the quasar fields to
search for faint objects near the quasar or the identified absorbing
galaxy, we cannot place more stringent limits on the definition
separating isolated and group galaxies. In the following section, we
present an averaged relationship between limiting absolute magnitude
and limiting apparent magnitude based upon the general specifications
of the published surveys (see Figures~\ref{fig:magvsz}a and
\ref{fig:magvsz}b). The curves show that the ``average'' survey is
generally not deep enough to extend down to luminous satellites like
the Large Magellanic Cloud around the Milky Way, and certainly cannot
detect dwarf galaxies in groups. Thus, the definition of isolated and
group galaxies refers, on average, to the brighter galaxy population.

Our ability to distinguish a group such as the Local Group based upon
its brightest members, M31 and the Milky Way, is, in the end,
dependent upon the direction from which the the Local Group would be
viewed. Given the $\sim 700$ kpc distance between M31 and the Milky
Way, a line of sight perpendicular to their separation vector would
render both galaxies as isolated regardless of their relative
sightline velocities. On the other hand, a line of sight more or less
parallel to their separation vector in which their projected
separation is less than 100 kpc would render their being classified as
group galaxies, given that their line of sight velocity separation is
on the order of $\sim 400$~{\kms}. As such, there is likely some
level, which is difficult to quantify, to which the isolated sample
contains galaxies in environments similar to the Local Group. And,
for those galaxies classified as group members, there may be some
fraction for which the galaxies are physically separated by a large
enough distance that their respective CGM environments do not overlap
into an intragroup medium (which is likely the case for M31 and the
Milky Way, but would in general depend upon the dynamical history of
the group).

As we stated above, we present only isolated galaxies in this paper
drawing upon a definition set by precedent in the literature. The
intragroup medium of the group galaxy subsample will be studied in a
forthcoming paper.

\section{Methods}
\label{sec:methods}

For each galaxy, the measured quantities include one or more apparent
magnitudes in various photometric bands, the spectroscopic redshift,
$z_{\rm gal}$, the galaxy right ascension and declination offsets from
the quasar, $\Delta\alpha$ and $\Delta\delta$, and/or the galaxy
angular separation from the quasar, $\theta$. Many of the galaxies
have multiple measurements from several different studies; we selected
the highest quality measurements (usually the most recent). 

\subsection{Galaxy Properties}
\label{sec:magcalcs}

Galaxy redshifts, $z_{\rm gal}$, were taken directly from published
values. The galaxy spectra were not published in most cases and were
not available in electronic form for further confirmation or
re-measurement. Since the uncertainties of $z_{\rm gal}$ were also not
published, the accuracy of the published $z_{\rm gal}$ measurements is
reflected by the number of significant figures.

Due to the application of different cosmologies in the literature over
the last $\sim 20$ years, we calculated new impact parameters, $D$,
and luminous properties for each galaxy using the $\Lambda$CDM
cosmology ($H_0=70$ km s$^{-1}$ Mpc$^{-1}$, $\Omega_M=0.3$, and
$\Omega_{\Lambda}=0.7$).

We first calculated the values $\Delta\alpha$ and $\Delta\delta$ from
the right ascensions and declinations of the galaxy and its associated
quasar, which were obtained from NED when available. We then
determined $\theta$ from $\Delta\alpha$ and $\Delta\delta$ using
$\theta=[(\Delta\delta)^2+(\Delta\alpha)^2~{\rm cos}^2(\delta_{\rm
    QSO})]^{1/2}$. We compute the impact parameter from $D=\theta
D_{\hbox{\tiny A}}(z_{\rm gal})$, where $D_{\hbox{\tiny A}}(z_{\rm
  gal})$ is the angular diameter distance at the galaxy redshift. In
cases where $\Delta\alpha$ and $\Delta\delta$ could not be obtained,
we used the published values of $\theta$ to calculate $D$. Using these
methods, we have measured $\Delta\alpha$ and $\Delta\delta$ for most
galaxies and standardized impact parameters for all galaxies in
{\magiicat}.

All galaxies have been imaged in the $g$, $r$, $R_{\hbox{\tiny
    EFOSC}}$, $R_s$, $R_{\hbox{\tiny J}}$, F702W, or F814W bands. For
each galaxy, we determined rest-frame AB absolute $B$-band magnitudes,
$M_B$, using the equation, $M_B({\rm AB})=[m_y-K_{By}]-DM$, where
$m_y$ is the AB apparent magnitude in the observed band, $K_{By}$ is
the appropriate $K$-correction \citep[e.g.,][see
  Appendix~\ref{app:kcorr}]{kim} for the observed magnitude, and $DM$
is the distance modulus for each galaxy. In cases where $m_y$ is a
Vega magnitude, such as F702W and F814W, we add the constant $-0.0873$
to convert from Vega magnitudes to AB magnitudes. Details on how we
determined this constant are given in Appendix~\ref{app:kcorr}.

To compute the $K$-corrections, we applied the actual filter response
curve for each published apparent magnitude. The $K$-corrections as a
function of redshift for the F702W- and $g$-band to the $B$-band are
shown in panels $a$ and $c$ of Figure~\ref{fig:kcorrs},
respectively. Using the spectral energy distribution (SED) templates
from \citet{bolzonella00} who extended the \citet{cww80} SEDs to
shorter and longer wavelengths, we adopted a SED for each galaxy. To
do this, we compared the observed color of each SED to each galaxy's
observed color and chose the SED with the closest color. For 18
galaxies where no observed galaxy color was available, we adopted an
Sbc galaxy SED, the average type selected by {\MgII} absorption
\citep[SDP94;][]{zibetti07}. Figure~\ref{fig:SEDcolors} in
Appendix~\ref{app:SEDcolors} presents ${\rm F702W}-K_s$ and $g-r$, the
two most common observed galaxy colors in {\magiicat}, as a function
of redshift.

The apparent magnitudes in {\magiicat} have not been Galactic
reddening corrected to the extent of our knowledge. We did not apply
this correction to the magnitudes as the mean reddening correction in
each magnitude band is, on average, a small fraction of the
uncertainty in the $K$-corrections due to SED selections. The greatest
reddening correction in {\magiicat} would be applied to the $g$ band,
as it is the band measuring the shortest wavelengths. The mean
reddening correction in this band is 0.2 magnitudes, while the
greatest difference in the $K$-correction $K_{B, g}$ between an E and
an Im SED is 1.5 magnitudes (see Figure~\ref{fig:kcorrs}).

Galaxies drawn from SDP94 and Steidel97 were imaged in the infrared
with the $K_s$ band. Many galaxies in CT08, BC09, KCBC10, KCEMS11, and
Churchill13 were imaged in either the 2MASS $K_s$ band and obtained
from NED, the $K'$ band from \citet{chen01b}, or the {\it HST} F160W
band from David Law (private communication). Using these values, we
computed $M_K({\rm AB})$ from $K$-corrected infrared magnitudes using
the methods applied for the $B$-band and the equation, $M_K({\rm
  AB})=[m_y-K_{Ky}]-DM+1.8266$, where $m_y$ is the Vega apparent
magnitude and the value $1.8266$ is the constant used to convert from
Vega magnitudes to AB magnitudes in all cases. The $K$-corrections in
the $K_s$ band for each SED are presented in panel $b$ of
Figure~\ref{fig:kcorrs} as a function of redshift.

\begin{figure*}[th]
\includegraphics[angle=0,trim = 3mm 0mm 0mm 0mm,scale=0.67]{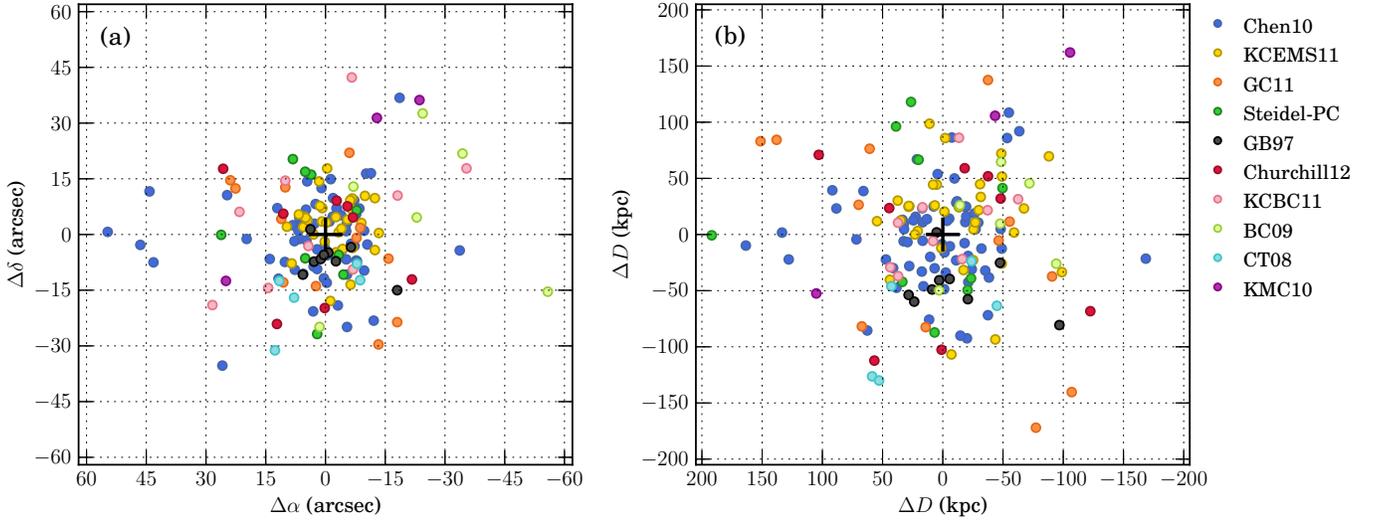}
\caption[]{($a$) Angular offsets (arcsec) of each galaxy for which we
  have $\Delta \alpha$ and $\Delta \delta$ from the associated
  background quasar. ($b$) Physical offsets (kpc) of each galaxy from
  the associated background quasar. Points are color coded by the work
  from which the galaxy was obtained. The plus sign indicates the
  location of the background quasar.}
\label{fig:radec}
\end{figure*}

\begin{figure}[th]
\includegraphics[angle=0,trim = 3mm 0mm 0mm 0mm,scale=0.57]{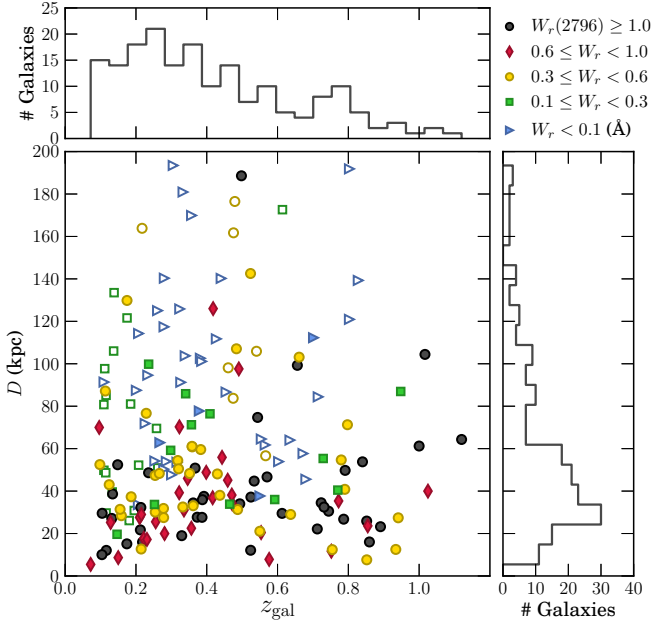}
\caption[]{The quasar-galaxy impact parameter, $D$, as a function of
  galaxy redshift, $z_{\rm gal}$. Points are colored by $W_r(2796)$,
  with open points representing $3~\sigma$ upper limits on
  $W_r(2796)$. Open yellow circles indicate galaxies with an upper
  limit of $W_r(2796)=0.3$~{\AA}, the largest upper limit in
  {\magiicat}. Histograms show the distribution of $D$ and $z_{\rm
    gal}$ for the full sample.}
\label{fig:Dvsz}
\end{figure}

\begin{figure}[th]
\includegraphics[angle=0,trim = 3mm 0mm 0mm 0mm,scale=0.57]{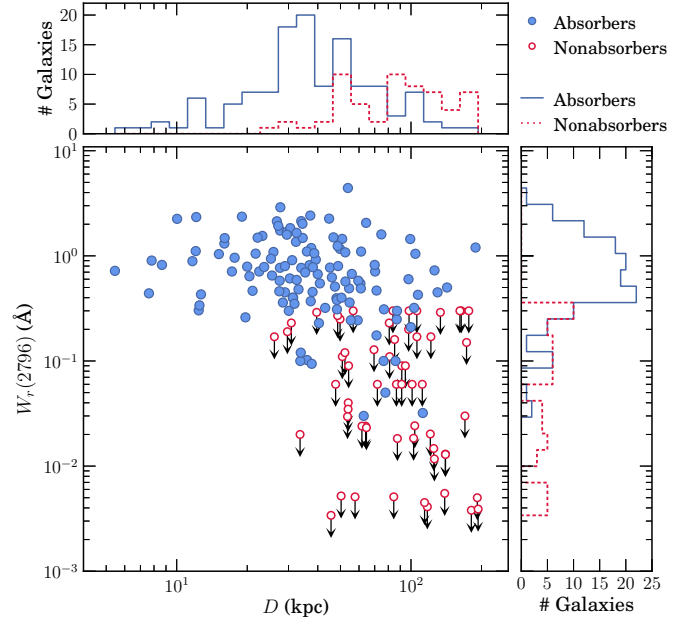}
\caption[]{The rest-frame {\MgII} equivalent width, $W_r(2796)$,
  versus impact parameter, $D$. Points are colored based on
  $W_r(2796)$ absorption, with solid blue points representing galaxies
  with measured absorption and red open points with downward arrows
  representing $3~\sigma$ upper limits on $W_r(2796)$. The histograms
  show the $D$ and $W_r(2796)$ distributions for absorbers (thin blue
  line) and an upper limit on absorption (dotted red line). The
  anti-correlation between $W_r(2796)$ and $D$ is significant at the
  $7.9~\sigma$ level.}
\label{fig:EWvsD}
\end{figure}

Apparent magnitudes in the $K$-band were not available for all
galaxies drawn from Chen10 and GC11, and many galaxies from BC09 and
KCBC11. We used an indirect method to compute $M_K$ by determining
rest-frame $B-K$ colors from rest-frame $B-R$ colors. We obtained
``model'' $g$ and $r$ apparent magnitudes from NED/SDSS. We adopt
``model'' magnitudes because the galaxy light is measured consistently
through the same aperture in all bands, therefore they are the best
magnitudes for measuring the colors of galaxies. Using these
magnitudes and the methods for $M_B$ above, we calculated $M_B$ from
$g$ and $M_R$ from $r$. The $K$-corrections for these magnitudes are
presented in panels $c$ and $d$ of Figure~\ref{fig:kcorrs},
respectively. In order to convert these $B-R$ colors to $B-K$, we
computed rest-frame $B-R$ and $B-K$ colors for each galaxy SED, which
suggest a linear relationship with the form $(B-K)=1.86(B-R)+0.02$,
determined from a linear least-squares fit to the rest-frame SED
colors. The rest-frame colors of each SED and the linear fit are
presented in Figure~\ref{fig:BKBR}. We then applied this relation to
the $B-R$ colors to obtain $B-K$. Finally, $M_K$ was calculated from
$M_B$ and $B-K$. Using these methods, we obtained colors for all but
18 galaxies in {\magiicat}.

\begin{deluxetable*}{lllrrrcccccccc}
\tabletypesize{\footnotesize}
\tablecolumns{15}
\tablewidth{0pt}
\setlength{\tabcolsep}{0.06in}
\tablecaption{Observed Galaxy Properties~\tablenotemark{a} \label{tab:obsprops}}
\tablehead{
  \colhead{}                         &
  \colhead{}                         &
  \colhead{}                         &
  \multicolumn{4}{c}{---------------- Galaxy ID ----------------} &  
  \multicolumn{3}{c}{------------ $B$-band ------------} &
  \multicolumn{3}{c}{---------- $K$-band ----------} &
  \colhead{} \\
  \colhead{(1)}                      &
  \colhead{(2)}                      &
  \colhead{(3)}                      &
  \colhead{(4)}                      &
  \colhead{(5)}                      &
  \colhead{(6)}                      &
  \colhead{(7)}                      &
  \colhead{(8)}                      &
  \colhead{(9)}                      &
  \colhead{(10)}                     &
  \colhead{(11)}                     &
  \colhead{(12)}                     &
  \colhead{(13)}                     &
  \colhead{(14)}                     \\
  \colhead{QSO}                      &
  \colhead{J-Name}                   &
  \colhead{$z_{\rm gal}$}              &
  \colhead{$\Delta\alpha$}           &
  \colhead{$\Delta\delta$}           &
  \colhead{$\theta$}                 &
  \colhead{Ref~\tablenotemark{b}}    &
  \colhead{$m_{y}$\tablenotemark{c}}  &
  \colhead{Band~\tablenotemark{d}}   &
  \colhead{Ref~\tablenotemark{b}}    &
  \colhead{$m_{y}$\tablenotemark{e}}  &
  \colhead{Band~\tablenotemark{d}}   &
  \colhead{Ref~\tablenotemark{b}}    &
  \colhead{SED~\tablenotemark{f}}    \\
  \colhead{}                         &
  \colhead{}                         &
  \colhead{}                         &
  \colhead{(arcsec)}                 &
  \colhead{(arcsec)}                 &
  \colhead{(arcsec)}                 &
  \colhead{}                         &
  \colhead{}                         &
  \colhead{}                         &
  \colhead{}                         &
  \colhead{}                         &
  \colhead{}                         &
  \colhead{}                         &
  \colhead{}                         }            
\startdata
0002$-$422  & J000448.11$-$415728.8  & $0.840   $ &   $-6.4$ &   $-3.4$ & $  7.10$ & 1  & $22.60$ & $R_{\hbox{\tiny{EFOSC}}}$(V) & 1   & $\cdots$ & $\cdots$ & $\cdots $ & (Sbc)    \\[2pt]
0002+051    & J000520.21+052411.80   & $0.298   $ &  $-13.4$ &    $0.4$ & $ 13.45$ & 3  & $19.86$ & F702W(V)                     & 3   &  $16.37$ & $K_s$(V) & $7      $ & E/S0     \\[2pt]
0002+051    & J000520.21+052411.80   & $0.592   $ &   $-2.6$ &   $-4.8$ & $  5.46$ & 3  & $21.11$ & F702W(V)                     & 3   &  $17.40$ & $K_s$(V) & $7      $ & E/S0     \\[2pt]
0002+051    & J000520.21+052411.80   & $0.85180 $ &   $-3.3$ &    $0.6$ & $  3.40$ & 3  & $22.21$ & F702W(V)                     & 3   &  $19.30$ & $K_s$(V) & $7      $ & Im       \\[2pt]
SDSS        & J003340.21$-$005525.53 & $0.2124  $ &   $-5.4$ &    $3.2$ & $  6.28$ & 6  & $19.44$ & $g$(AB)                      & 14  &  $18.79$ & $r$(AB)  & $14     $ & Scd      \\[2pt]
SDSS        & J003407.34$-$085452.07 & $0.3617  $ &    $6.5$ &   $-1.2$ & $  6.56$ & 6  & $22.41$ & $g$(AB)                      & 14  &  $21.45$ & $r$(AB)  & $14     $ & Scd      \\[2pt]
SDSS        & J003413.04$-$010026.86 & $0.2564  $ &   $-2.8$ &    $7.1$ & $  7.63$ & 6  & $21.68$ & $g$(AB)                      & 14  &  $20.25$ & $r$(AB)  & $14     $ & E/S0     \\[2pt]
0058+019    & J010054.15+021136.52   & $0.6128  $ & $\cdots$ & $\cdots$ & $  4.40$ & 7  & $23.25$ & $R_s$(AB)                    & 7   &  $19.90$ & $K_s$(V) & $7      $ & Sbc      \\[2pt]
0058+019    & J010054.15+021136.52   & $0.680   $ &   $-3.3$ &   $-5.5$ & $  6.50$ & 2  & $22.06$ & F702W(V)                     & 2   &  $18.65$ & $K_s$(V) & $8      $ & Sbc      \\[2pt]
SDSS        & J010135.84$-$005009.08 & $0.2615  $ &   $10.2$ &   $-7.4$ & $ 12.60$ & 6  & $20.91$ & $g$(AB)                      & 14  &  $19.57$ & $r$(AB)  & $14     $ & E/S0     \\[-5pt]
\enddata
\tablenotetext{a}{Table~\ref{tab:obsprops} is published in its entirety
  in the electronic edition of ApJ. A portion is shown here for
  guidance regarding its form and content.}
\tablenotetext{b}{Galaxy Identification and Apparent Magnitude
  Reference: (1) \citet{guillemin}, (2) \citet{churchill-weakgals},
  (3) \citet{kcems}, (4) \citet{steidel97}, (5) \citet{chen08}, (6)
  \citet{chen10a}, (7) \citet{sdp94}, (8) Steidel-PC, (9)
  \citet{ggk1127}, (10) \citet{kacprzak11kin}, (11)
  \citet{gauthier11}, (12) \citet{barton09}, (13) \citet{chen01b},
  (14) NED/SDSS, (15) NED/2MASS, and (16) David Law, personal
  communication.}
\tablenotetext{c}{Apparent magnitude used to obtain $M_{B}$.}
\tablenotetext{d}{Magnitude Band and Type: (AB) AB magnitude, and (V)
  Vega magnitude.}
\tablenotetext{e}{Apparent magnitude used to obtain $M_K$.}
\tablenotetext{f}{Galaxy Spectral Energy Distributions: (Sbc) No color
  information -- Sbc used.}
\end{deluxetable*}

$B$-band luminosities, $L_B/L_B^{\ast}$, were obtained using a linear
fit to $M_B^{\ast}$ with redshift using the ``All'' sample in Table~6
from \citet{faber}. The $K$-band luminosities, $L_K/L_K^{\ast}$, were
computed using $M_K^{\ast}(z)$ as expressed in Eq.~2 from
\citet{cirasuolo}.

\subsection{Absorption Properties}

Where we have obtained access to HIRES/Keck or UVES/VLT quasar
spectra, we have remeasured $W_r(2796)$ using the methods of
\citet{schneider93} and \citet{archiveI}. Upper limits on $W_r(2796)$
are quoted at $3~\sigma$ and must be less than or equal to 0.3~{\AA},
corresponding to an unresolved absorption feature, for a galaxy to be
included in {\magiicat}. In cases where HIRES/Keck and/or UVES/VLT
spectra do not exist, we adopted the best published values. In most
cases, these are measurements are take directly from tabulated data.
We have also converted published $2~\sigma$ upper limits to $3~\sigma$
where needed. We have not included galaxies with upper limits less
stringent than $0.3$~{\AA} in order to ensure all non-detections
reside below the historical absorption threshold of 0.3~{\AA} (SDP94).

Adopting a $3~\sigma$ upper limit of $W_r(2796)=0.3$~{\AA} allows us
to distinguish two subsamples by absorption strength based upon
historical precedence. Several of the surveys from which {\magiicat}
is constructed were conducted on 4-meter class telescopes using
moderate resolution ($\sim 1$~{\AA}); the typical equivalent width
detection sensitivity of these surveys was $W_r(2796)=0.3$~{\AA}. Not
until the advent of the HIRES spectrograph on Keck~I in the early
1990s was it possible to systematically explore equivalent widths
below this threshold \citep{weakI}. Absorbers with $W_r(2796)<
0.3$~{\AA} were thus dubbed ``weak systems''. 

Following the historical precedent, we adopt the term ``strong''
absorption for values in the range $W_r(2796) \geq 0.3$~{\AA}.
However, we note that this term has often been applied in the
literature to describe absorbers with $W_r(2796) \geq 1.0$~{\AA} or
higher. Throughout this work, we also adopt the term ``weak''
absorption for values in the range $W_r(2796)< 0.3$~{\AA}, regardless
of whether absorption was formally detected or not detected to the
varying limits of the HIRES and/or UVES quasar spectra.

The term ``non-absorber'' is most suitably applied for surveys
designed to be complete to a well-defined detection threshold, in
which upper limits below the threshold would be described using such a
term to indicate ``below the survey limit''. In a global description
of absorption strengths, we must keep in mind that a non-detection
does not provide evidence for no absorption. A highly stringent limit
on $W_r(2796)$ could indicate that the line of sight passed through a
low-density, high-ionization region in which the ionization fraction
of {\MgII} is vanishingly small. Similarly, it could indicate that
the line of sight probed a very low {\HI} column density, dense,
low-ionization structure, such that even with a high metallicity and a
high ionization fraction, {\MgII} would not be detected to the limits
of the data. It is likely extremely rare that gas within 200 kpc
(projected) of a galaxy and relative velocity within a few hundred
kilometers per second would be entirely devoid of {\MgII} absorption.

In order to simplify terminology, we adopt the two terms ``weak'' and
``strong'' absorption, and remind the reader that ``weak'' absorption
encompasses all measurements below $W_r(2796) = 0.3$~{\AA} and all
upper limits, since presumably higher and higher sensitivities would
yield ``detections'' in most all cases.

\section{Galaxy Sample}
\label{sec:sample}

Here we present basic characteristics of {\magiicat}, leaving any
analysis to future work.

~{\magiicat} consists of 182 isolated galaxies along 134
sightlines. The redshift range of the sample is $0.07 \leq z \leq
1.12$, with median $\langle z \rangle = 0.359$. 

\begin{figure*}[th]
\includegraphics[angle=0,trim = 3mm 0mm 0mm 0mm,scale=0.57]{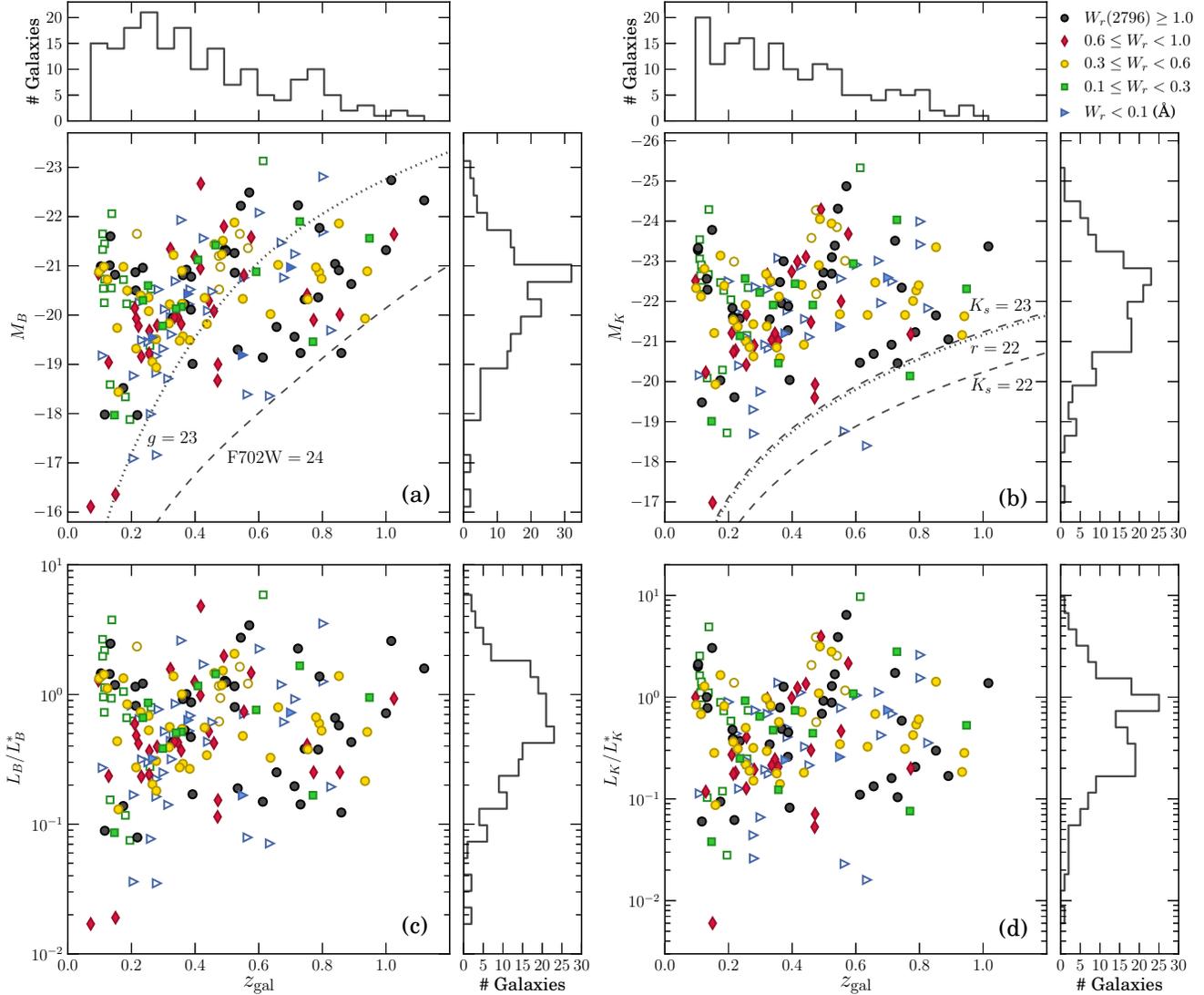}
\caption[]{Galaxy luminous properties, $M_B$, $M_K$, $L_B/L_B^{\ast}$,
  and $L_K/L_K^{\ast}$, as a function of redshift. Point types
  indicate the strength of {\MgII} absorption, $W_r(2796)$. Open
  points represent upper limits on absorption. The distributions of
  redshift and luminous properties are shown in histograms along the
  respective axes. ($a$) The $B$-band AB absolute magnitude, $M_B$, as
  a function of galaxy redshift, $z_{\rm gal}$. $M_B$ is calculated at
  all redshifts from apparent magnitudes of $g=23$ (AB magnitude;
  dotted line) and ${\rm F702W}=24$ (Vega magnitude; dashed line),
  representing the limiting magnitudes for the surveys in which many
  of the galaxies were observed. The value of $g$ comes from the Sloan
  Digital Sky Survey, whereas the F702W-band was observed with WFPC2
  on {\it HST}. ($b$) The $K$-band AB absolute magnitude, $M_K$,
  versus galaxy redshift, $z_{\rm gal}$. $M_K$ is calculated at all
  redshifts from apparent $K_s$- and $r$-band magnitudes to indicate
  the limiting magnitudes associated with the various samples in
  {\magiicat}. The majority of galaxies imaged with the $K_s$-band
  have a limiting magnitude of $K_s=21$ (Vega magnitude; dashed line),
  while those obtained from Steidel97 were imaged more deeply down to
  $K_s=22$ (Vega magnitude; dashed line). Galaxies imaged in the
  $r$-band with SDSS have a limiting magnitude of $r=22$ (dotted
  line). We translated $M_r$ into $M_K$ using the relationship between
  $B-R$ and $B-K$ colors in \S~\ref{sec:magcalcs} and
  Appendix~\ref{app:BKBR}. ($c$) The $B$-band luminosity,
  $L_B/L_B^{\ast}$, as a function of galaxy redshift, $z_{\rm
    gal}$. ($d$) The $K$-band luminosity, $L_K/L_K^{\ast}$, versus
  galaxy redshift, $z_{\rm gal}$.}
\label{fig:magvsz}
\end{figure*}

Observed galaxy properties for {\magiicat} are presented in
Table~\ref{tab:obsprops}. The columns include the (1) QSO identifier,
(2) Julian 2000 designation (J-Name), (3) galaxy spectroscopic redshift,
$z_{\rm gal}$, (4) quasar-galaxy right ascension offset,
$\Delta\alpha$, (5) quasar-galaxy declination offset, $\Delta\delta$,
(6) quasar-galaxy angular separation, $\theta$, (7) reference for
columns 4, 5, and 6, (8) apparent magnitude used to obtain $M_B$, (9)
band for the preceding apparent magnitude, (10) reference for columns
8 and 9, (11) apparent magnitude used to calculate $M_K$, (12) band
for $m_K$, (13) reference for columns 11 and 12, and (14) galaxy
spectral energy distribution type based on the galaxy observed color.

The galaxy right ascension and declination offsets from the quasar,
$\Delta\alpha$ and $\Delta\delta$, are presented in
Figure~\ref{fig:radec}$a$. Points are colored by reference, indicating
the work from which the galaxy data was drawn. No values for
$\Delta\alpha$ and $\Delta\delta$ were originally published for
galaxies from SDP94, though later works have obtained the values for
many of these galaxies (Steidel, private communication). The plus sign
indicates the location of the associated background quasar.

Figure~\ref{fig:radec}$b$ shows the location of each galaxy in
{\magiicat} in physical units (kpc) with respect to the associated
background quasar (plus sign). The points are colored by the source of
the $\Delta\alpha$ and $\Delta\delta$ measurements.

Impact parameter, $D$, as a function of galaxy redshift, $z_{\rm
  gal}$, is presented in Figure~\ref{fig:Dvsz}. Points are colored by
$W_r(2796)$, with $3~\sigma$ upper limits on $W_r(2796)$ represented
as open points. Histograms of the data collapsed along the axes show
the distribution of impact parameters and galaxy redshifts. Impact
parameters range from $5.4 \leq D \leq 194$~kpc, where the median
impact parameter is $\langle D \rangle = 48.7$~kpc.

\begin{turnpage}
\begin{deluxetable*}{llllcccrrrcrccc}
\tabletypesize{\footnotesize}
\tablecolumns{15}
\tablewidth{0pt}
\setlength{\tabcolsep}{0.06in}
\tablecaption{Calculated Galaxy and Absorption Properties~\tablenotemark{a} \label{tab:calcprops}}
\tablehead{
  \colhead{}                         &
  \colhead{}                         &
  \colhead{}                         &
  \multicolumn{4}{c}{-------------------- {\MgII} Absorption --------------------} &  
  \colhead{}                         &
  \multicolumn{3}{c}{---------- $B$-band ----------} &
  \multicolumn{3}{c}{---------- $K$-band ----------} &
  \colhead{} \\
  \colhead{(1)}                      &
  \colhead{(2)}                      &
  \colhead{(3)}                      &
  \colhead{(4)}                      &
  \colhead{(5)}                      &
  \colhead{(6)}                      &
  \colhead{(7)}                      &
  \colhead{(8)}                      &
  \colhead{(9)}                      &
  \colhead{(10)}                     &
  \colhead{(11)}                     &
  \colhead{(12)}                     &
  \colhead{(13)}                     &
  \colhead{(14)}                     &
  \colhead{(15)}                     \\
  \colhead{QSO}                      &
  \colhead{J-Name}                   &
  \colhead{$z_{\rm gal}$}              &
  \colhead{$z_{\rm abs}$}              &
  \colhead{$W_r(2796)$}              &
  \colhead{$DR$}                     &
  \colhead{Ref~\tablenotemark{b}}    &
  \colhead{$D$}                      &
  \colhead{$K_{By}$\tablenotemark{c}} &
  \colhead{$M_B$\tablenotemark{d}}   &
  \colhead{$L_B/L_B^{\ast}$}           &
  \colhead{$K_{Ky}$\tablenotemark{e}} &
  \colhead{$M_K$\tablenotemark{d}}   &
  \colhead{$L_K/L_K^{\ast}$}           &
  \colhead{$B-K$}                    \\
  \colhead{}                         &
  \colhead{}                         &
  \colhead{}                         &
  \colhead{}                         &
  \colhead{\AA}                      &
  \colhead{}                         &
  \colhead{}                         &
  \colhead{(kpc)}                    &
  \colhead{}                         &
  \colhead{}                         &
  \colhead{}                         &
  \colhead{}                         &
  \colhead{}                         &
  \colhead{}                         &
  \colhead{}                         }            
\startdata
0002$-$422  & J000448.11$-$415728.8  & $0.840   $ & $0.836627$ & $4.422\pm0.002$ & $1.12\pm0.09$   & 13 & $  53.8$ & $-0.08$ & $-21.04$ & $0.66$ & $\cdots$ $$ & $ \cdots$ & $\cdots$ & $\cdots$ \\[2pt]
0002+051    & J000520.21+052411.80   & $0.298   $ & $0.298059$ & $0.244\pm0.003$ & $1.336\pm0.029$ & 3  & $  59.2$ & $-1.39$ & $-19.78$ & $0.38$ & $    -0.52$ & $ -22.22$ & $  0.64$ & $2.43  $ \\[2pt]
0002+051    & J000520.21+052411.80   & $0.592   $ & $0.591365$ & $0.102\pm0.002$ & $1.539\pm0.039$ & 3  & $  36.0$ & $-0.80$ & $-20.88$ & $0.76$ & $    -0.53$ & $ -22.94$ & $  1.08$ & $2.05  $ \\[2pt]
0002+051    & J000520.21+052411.80   & $0.85180 $ & $0.851393$ & $1.089\pm0.008$ & $1.160\pm0.013$ & 3  & $  25.9$ & $-0.64$ & $-20.91$ & $0.58$ & $    -0.88$ & $ -21.65$ & $  0.29$ & $0.74  $ \\[2pt]
SDSS        & J003340.21$-$005525.53 & $0.2124  $ & $0.2121  $ & $1.05\pm0.03$   & $\cdots$        & 6  & $  21.7$ & $ 0.20$ & $-20.87$ & $1.15$ & $     0.06$ & $ -21.83$ & $  0.47$ & $0.96  $ \\[2pt]
SDSS        & J003407.34$-$085452.07 & $0.3617  $ & $0.3616  $ & $0.48\pm0.05$   & $\cdots$        & 6  & $  33.1$ & $ 0.54$ & $-19.57$ & $0.29$ & $     0.13$ & $ -20.59$ & $  0.14$ & $1.02  $ \\[2pt]
SDSS        & J003413.04$-$010026.86 & $0.2564  $ & $0.2564  $ & $0.61\pm0.06$   & $\cdots$        & 6  & $  30.4$ & $ 0.80$ & $-19.68$ & $0.37$ & $     0.43$ & $ -21.68$ & $  0.40$ & $1.99  $ \\[2pt]
0058+019    & J010054.15+021136.52   & $0.6128  $ & $0.612586$ & $1.684\pm0.004$ & $1.06\pm0.09$   & 13 & $  29.5$ & $-0.40$ & $-19.14$ & $0.15$ & $    -0.59$ & $ -20.47$ & $  0.11$ & $1.32  $ \\[2pt]
0058+019    & J010054.15+021136.52   & $0.680   $ & $0.680   $ & $<0.0034$       & $\cdots$        & 2  & $  45.6$ & $-0.34$ & $-20.76$ & $0.61$ & $    -0.63$ & $ -21.96$ & $  0.42$ & $1.20  $ \\[2pt]
SDSS        & J010135.84$-$005009.08 & $0.2615  $ & $0.2615  $ & $<0.11$         & $\cdots$        & 6  & $  50.9$ & $ 0.82$ & $-20.53$ & $0.80$ & $     0.44$ & $ -22.34$ & $  0.74$ & $1.81  $ \\[-5pt]
\enddata
\tablenotetext{a}{Table~\ref{tab:calcprops} is published in its entirety
  in the electronic edition of ApJ. A portion is shown here for
  guidance regarding its form and content.}
\tablenotetext{b}{{\MgII} Absorption Measurements: (1)
  \citet{guillemin}, (2) \citet{churchill-weakgals}, (3)
  \citet{kcems}, (4) \citet{steidel97}, (5) \citet{chen08}, (6)
  \citet{chen10a}, (7) \citet{sdp94}, (8) Steidel-PC, (9)
  \citet{ggk1127}, (10) \citet{kacprzak11kin}, (11)
  \citet{gauthier11}, (12) \citet{barton09}, (13)
  \citet{evans-thesis}, and (14) This work.}
\tablenotetext{c}{$K$-correction used to obtain $M_B$ from column (8)
  in Table~\ref{tab:obsprops}.}
\tablenotetext{d}{Absolute magnitudes are AB magnitudes.}
\tablenotetext{e}{$K$-correction used to obtain $M_K$ from column (11)
  in Table~\ref{tab:obsprops}.}
\end{deluxetable*}
\end{turnpage}

\begin{figure}[th]
\includegraphics[angle=0,trim = 3mm 0mm 0mm 0mm,scale=0.58]{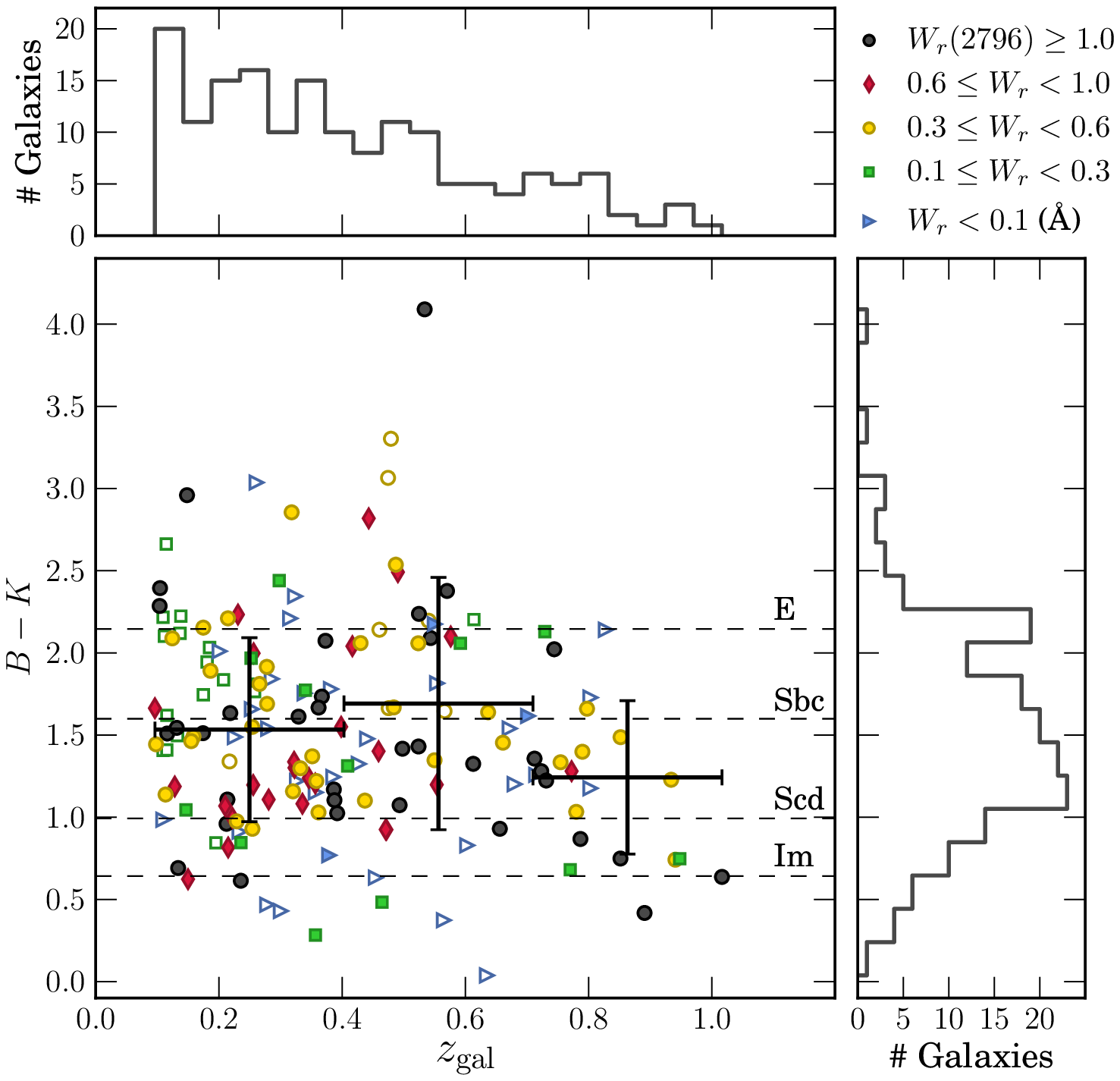}
\caption[]{Rest-frame galaxy color, $B-K$, against galaxy redshift,
  $z_{\rm gal}$. Points are colored by $W_r(2796)$, with open points
  representing upper limits on $W_r(2796)$. The dashed lines mark the
  rest-frame $B-K$ colors for the spectral energy distributions of E,
  Sbc, Scd, and Im galaxies. Crosses indicate the average color in
  three redshift bins ($z<0.403$, $0.403 \leq z < 0.709$, and $z \geq
  0.709$). Horizontal error bars show the range in redshift for each
  bin while vertical error bars are the standard deviations in $B-K$
  for each bin. The average colors in all three redshift bins are
  consistent with an Sbc galaxy. Color and redshift distributions for
  {\magiicat} are presented as histograms. From the color histogram,
  the most common galaxy type is slightly redder than an Scd galaxy.}
\label{fig:BKvsz}
\end{figure}

From Figure~\ref{fig:Dvsz}, it is apparent that most galaxies with
upper limits on absorption are found at larger impact parameters. This
anti-correlation between $W_r(2796)$ and $D$ is a commonly known
property of {\MgII} galaxies \citep[e.g.,][]{lanzetta90, bb91,
  steidel95, bouche06, ggk08, chen10a, churchill-weakgals}. We
performed a non-parametric Kendall's $\tau$ rank correlation test on
$W_r(2796)$ and $D$, allowing for upper limits on $W_r(2796)$
\citep[see][]{bhk,wang}. We find that $W_r(2796)$ is anti-correlated
with $D$ at the $7.9~\sigma$ level. We present the anti-correlation in
Figure~\ref{fig:EWvsD}, leaving any analysis for future work in which
we explore the scatter in the relationship as a function of galaxy
properties \citep[see Paper II and][]{churchill-masses}. Galaxies with
measured absorption are shown as blue points, while upper limits on
absorption are open red circles with downward arrows. The
distributions of $W_r(2796)$ and $D$ for absorbers (thin blue lines),
and galaxies with upper limits on absorption (dotted red lines) are
presented in histograms.

The equivalent width detections and upper limits on absorption of
{\magiicat} galaxies have a range of $0.003 \leq W_r(2796) \leq
4.42$~{\AA}, where the weakest confirmed absorption is
$W_r(2796)=0.03$~{\AA}. Of the 182 galaxies, 59 have upper limits on
$W_r(2796)$ where the most stringent upper limit is 3~m{\AA}. To a
$3~\sigma$ $W_r(2796)$ threshold, {\magiicat} is 100\% complete to
0.3~{\AA}, 90\% complete to 0.2~{\AA}, 80\% complete to 0.05~{\AA},
and 70\% complete to 0.01~{\AA}.

The absolute $B$-band magnitudes range from $-16.1 \geq M_B \geq
-23.1$. Figure~\ref{fig:magvsz}$a$ shows $M_B$ as a function of
$z_{\rm gal}$. Points are colored by $W_r(2796)$, with open points
representing upper limits on $W_r(2796)$. Histograms show the
distribution of galaxies in redshift and $M_B$.

Absolute $K$-band magnitudes range from $-17.0 \geq M_K \geq
-25.3$. Figure~\ref{fig:magvsz}$b$ presents $M_K$ against $z_{\rm
  gal}$. Points are colored by $W_r(2796)$, with upper limits on
$W_r(2796)$ shown as open points. Histograms present the distribution
of redshift and $M_K$ of the sample.

\begin{deluxetable}{lcccc}
\tabletypesize{\footnotesize}
\tablecolumns{15}
\tablewidth{0pt}
\setlength{\tabcolsep}{0.06in}
\tablecaption{~{\magiicat} Properties \label{tab:catprops}}
\tablehead{
  \colhead{Property}                 &
  \colhead{Min}                      &
  \colhead{Max}                      &
  \colhead{Mean}                     &
  \colhead{Median}                   }
\startdata
$W_r(2796)$~({\AA}) & 0.003 & 4.422 & 0.629 & 0.400 \\[3pt]
$z_{\rm gal}$         & 0.072 & 1.120 & 0.418 & 0.359 \\[3pt]
$D$~(kpc)           &   5.4 & 193.5 &  61.1 &  48.7 \\[3pt]
$M_B$               & -16.1 & -23.1 & -20.3 & -20.4 \\[3pt]
$M_K$               & -17.0 & -25.3 & -21.9 & -22.0 \\[3pt]
$L_B/L_B^{\ast}$      & 0.017 & 5.869 & 0.855 & 0.611 \\[3pt]
$L_K/L_K^{\ast}$      & 0.006 & 9.712 & 0.883 & 0.493 \\[3pt]
$B-K$               &  0.04 &  4.09 &  1.54 &  1.48 \\[-5pt]
\enddata
\end{deluxetable}

The completeness limits of the magnitudes are complicated due to the
heterogeneous imaging campaigns described in
\S~\ref{sec:overview}. For the $B$-band, the majority of galaxies at
low redshift ($z < \langle z \rangle$, where $\langle z \rangle
=0.359$ is the median redshift of the sample) were selected by SDSS
$r$ magnitudes with a limiting magnitude of $r=22$. However, we used
the SDSS $g$ band for these galaxies to calculate $M_B$, which is
sensitive to the threshold $g \simeq 23$ (AB magnitude; dotted line in
Figure~\ref{fig:magvsz}$a$). At the mean redshift of the low redshift
subsample, $z=0.23$, this corresponds to $M_B \simeq -17.5$. The
majority of galaxies at high redshift ($z \geq \langle z \rangle$)
were imaged in the F702W-band. This subsample has a threshold of ${\rm
  F702W} \simeq 24$ (Vega magnitude; dashed line in
Figure~\ref{fig:magvsz}$a$), corresponding to $M_B \simeq -18$ at the
mean redshift, $z=0.61$.

In the infrared, the galaxy magnitudes are generally sensitive to the
threshold $K_s \simeq 21$ (Vega magnitude), with the exception of the
extensive campaign on the field 1622+268 (Steidel97) which is
sensitive to $K_s \simeq 22$ (Vega magnitude; dashed lines in
Figure~\ref{fig:magvsz}$b$). For the high redshift subsample, $K_s
\simeq 21$ corresponds $M_K \simeq -20$ at the mean redshift,
$z=0.61$. The majority of galaxies at low redshift were imaged in the
$r$-band with SDSS. This subsample has a threshold of $r \simeq 22$
(AB magnitude; dotted line in Figure~\ref{fig:magvsz}$b$), which, when
converted to $M_K$ using the conversion between $B-R$ and $B-K$, is
roughly equal to $K_s \simeq 21$.

$B$-band luminosities have a range of $0.02 \leq L_B/L_B^{\ast} \leq
5.87$, with median $L_B/L_B^{\ast}=0.611$. $L_B/L_B^{\ast}$ as a
function of $z_{\rm gal}$ is presented in
Figure~\ref{fig:magvsz}$c$. Point colors represent $W_r(2796)$, with
open points indicating upper limits on $W_r(2796)$. Histograms show
the distributions of $B$-band luminosity and galaxy redshift.

Luminosities in the $K$-band range from $0.006 \leq L_K/L_K^{\ast}
\leq 9.71$ and have a median of $L_K/L_K^{\ast}=0.493$. $K$-band
Luminosity as a function of $z_{\rm gal}$ is presented in
Figure~\ref{fig:magvsz}$d$ with point colors indicating $W_r(2796)$
and upper limits on $W_r(2796)$ as open points. The distributions of
$L_K/L_K^{\ast}$ and $z_{\rm gal}$ are presented in histograms along
their respective axes.

Galaxy rest-frame $B-K$ colors have a range of $0.04 \leq B-K \leq
4.09$, with median $B-K=1.48$. $B-K$ as a function of $z_{\rm gal}$ is shown in
Figure~\ref{fig:BKvsz} and point colors represent $W_r(2796)$
strength, with open points indicating upper limits on
$W_r(2796)$. Color and redshift distributions are shown in
histograms. The mean and standard deviations in $B-K$ for three
equal-sized redshift bins ($z<0.403$, $0.403 \leq z < 0.709$, and $z
\geq 0.709$) are plotted as black error bars. Horizontal error bars
indicate the range in redshift for each bin. The horizontal dashed
lines indicate the rest-frame $B-K$ color for each SED, where all
three redshift bins are consistent with an Sbc galaxy. This is in
agreement with SDP94 and \citet{zibetti07}, who find that galaxies
with {\MgII} absorption have, on average, an Sbc SED type. The most
common galaxy color is slightly redder than an Scd SED type.

Table~\ref{tab:calcprops} presents calculated galaxy and absorption
properties for galaxies in {\magiicat}. The listed columns are the (1)
QSO identifier, (2) Julian 2000 designation (J-Name), (3) galaxy
spectroscopic redshift, $z_{\rm gal}$, (4) {\MgII} absorption
redshift, $z_{\rm abs}$, (5) {\MgII} equivalent width, $W_r(2796)$,
(6) {\MgII} doublet ratio, (7) reference for columns 4, 5, and 6, (8)
quasar-galaxy impact parameter, $D$, (9) $K$-correction to obtain
$M_B$, (10) absolute $B$-band magnitude, $M_B$, (11) $B$-band
luminosity, $L_B/L_B^{\ast}$, (12) $K$-correction to obtain $M_K$,
(13) absolute $K$-band magnitude, $M_K$, (14) $K$-band luminosity,
$L_K/L_K^{\ast}$, and (15) rest-frame color, $B-K$.

A summary of the absorption and galaxy properties of {\magiicat} is
presented in Table~\ref{tab:catprops}. We list the minimum, maximum,
mean, and median values for each property.

\subsection{Luminosity Functions}
\label{sec:phiM}

Prior to measurements of galaxy luminosity functions, $\Phi(M)$, out
to $z=1$ \citep[e.g.,][]{lilly95, lin99, fried01, wolf03, faber,
  cirasuolo}, selecting galaxies by {\MgII} absorption provided a
compelling technique for compiling a presumably complete sample of
intermediate redshift galaxies SDP94. Using absorption selection,
SDP94 presented the $B$- and $K$-band luminosity functions of
intermediate redshift galaxies associated with $W_r(2796) \geq
0.3$~{\AA}.

In order to compare to and expand upon the work of SDP94, we measured
the $B$- and $K$-band luminosity functions for {\magiicat}
galaxies. We divided the galaxies at all redshifts into two subsamples
bifurcated by $W_r(2796)=0.3$~{\AA}. We also divided the galaxies into
four subsamples bifurcated by $W_r(2796)=0.3$~{\AA} and $z = 0.359$
(the median redshift). The average redshift is $z=0.23$ for the low
redshift subsample and $z=0.61$ for the high redshift subsample. These
averages translate to a $3.2$~Gyr time spread.

Since the majority of the galaxies are absorption selected, we
followed SDP94 and applied a gas cross section correction, $\left<
R_{\rm gas} \right> ^{-2}$, to the number of galaxies in each
luminosity bin, where $\left< R_{\rm gas} \right> \propto \left<
L/L^{\ast} \right> ^{\beta}$, where $\left< L/L^{\ast} \right>$ is the
mean luminosity of the bin and where $\beta$ gives the empirically
determined luminosity dependence. For the {\magiicat} galaxies, we
determined $\beta=0.38$ for the $B$-band, and $\beta=0.27$ for the
$K$-band, as described in Paper II \citep{nielsen12}. The correction
factor rectifies the relative volume probed by absorption line surveys
at fixed luminosity under the assumption of completeness.

In Figures~\ref{fig:Blumfunc} and \ref{fig:Klumfunc}, we present the
$B$- and $K$-band luminosity functions for {\magiicat} galaxies. For
reference, we have overplotted empirically determined Schechter
luminosity functions from deep galaxy surveys (plus a single additive
constant to roughly match the data at $M_{\ast}$). We show the $z=0.3$
(solid curve) and $z=1.1$ (dashed curve) luminosity functions from
\citet{faber} for the $B$-band using their ``all galaxy sample'' and
\citet{cirasuolo} for the $K$-band of galaxies in the UKIDSS UDF
field. These curves roughly bracket the low and high redshift
subsamples. For the $B$-band, the appropriate characteristic
luminosities are $M^{\ast}_B = -21.1$ for $z=0.3$ and $M^{\ast}_B =
-21.5$ for $z=1.1$. For the $K$-band, the appropriate characteristic
luminosities are $M^{\ast}_K = -22.7$ for $z=0.3$ and $M^{\ast}_K =
-23.1$ for $z=1.1$.

For each binned data point, we computed the mean $B-K$ rest-frame color
of the galaxies contributing to the bin and then color coded the point
based upon the closest matching SED type (see Figure~\ref{fig:BKvsz}).
Red data points indicate an average SED type for an elliptical (E)
galaxy, yellow indicates Sbc on average, green indicates Scd on
average, and blue indicates Magellanic-type irregular (Im) on average.

If we adopt the view that the galaxy surveys are magnitude-limited, we
estimate that the completeness of {\magiicat} begins to decline for
$M_B > -18$ and $M_K > -17.5$ for the low redshift subsample and $M_B
> -19$ and $M_K > -20$ for the high redshift subsample (see
Figures~\ref{fig:magvsz}$a$ and \ref{fig:magvsz}$b$). We
conservatively plotted the data in these luminosity bins as open
points.

\subsubsection{$W_r(2796)$ and Redshift Differences}

\begin{figure}[thb]
\includegraphics[trim = 5mm 0mm 0mm 0mm,scale=0.61]{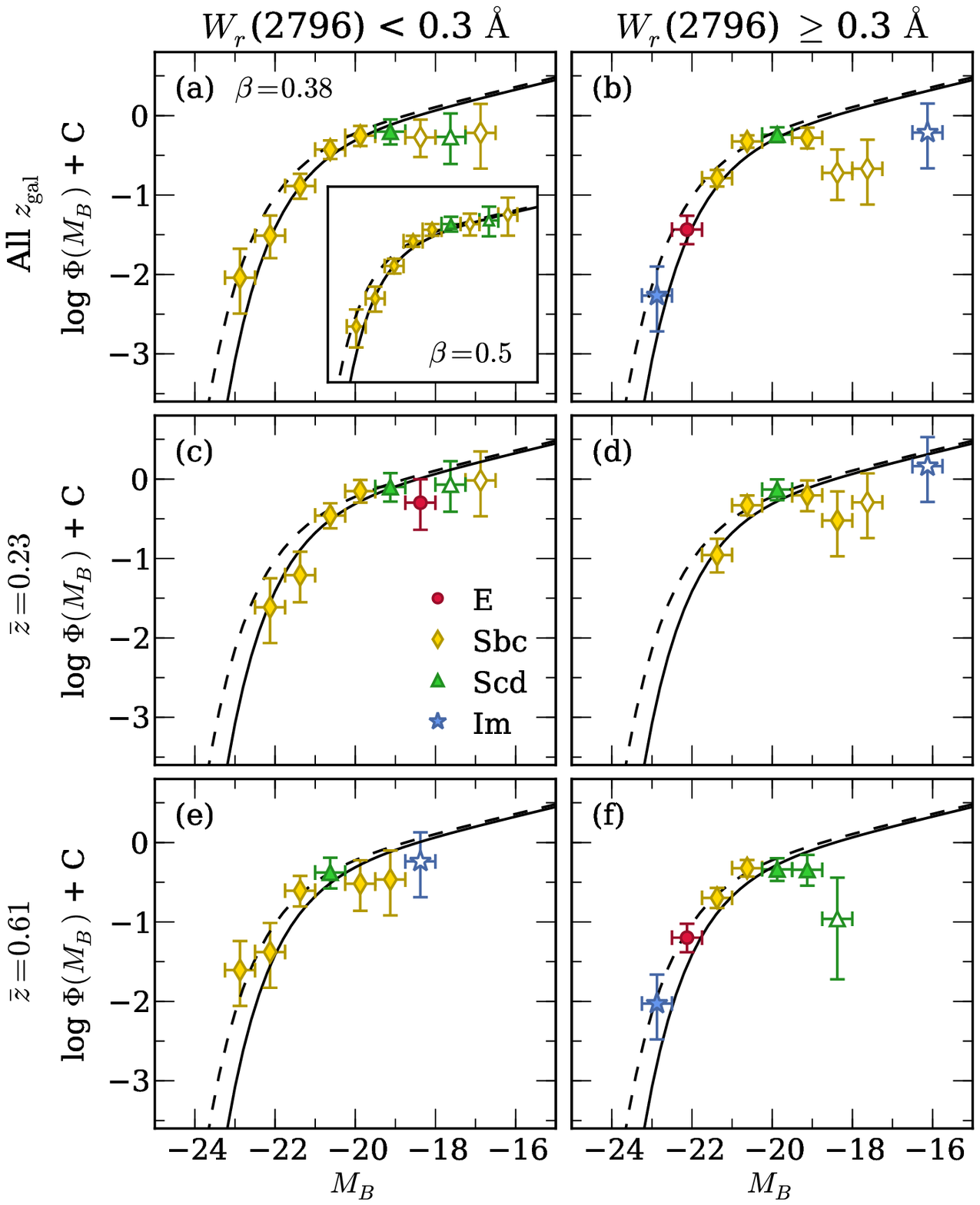}
\caption[]{$B$-band luminosity functions for $W_r(2796) < 0.3$ ($a$,
  $c$, and $e$) and $W_r(2796) \geq 0.3$ ($b$, $d$, and
  $f\,$). Galaxies at all redshifts are included in panels $a$ and
  $b$, while galaxies with $z < \langle z \rangle$ are in panels $c$
  and $d$ (mean $z=0.22$), and galaxies with $z \geq \langle z
  \rangle$ are in panels $e$ and $f$ (mean $z=0.61$), where $\langle z
  \rangle = 0.359$. The solid and dashed curves are $z = 0.3, 1.1$
  Schechter functions, respectively, from \citet{faber}. Data point
  colors and types are assigned according to the mean color of the
  galaxies in the bin. Red circles indicate the mean color in the bin
  is closest to an elliptical (E) SED type, yellow diamonds indicate
  Sbc, green triangles indicate Scd, and blue stars indicates a
  Magellanic-type irregular (Im) SED type. Open points are where the
  completeness of the sample declines. The inset of panel ($a$)
  illustrates how the observed luminosity function changes with the
  gas cross section correction $\langle R_{\rm gas} \rangle ^{-2} \propto
  \langle L/L_B^{\ast} \rangle^{-2\beta}$.}
\label{fig:Blumfunc}
\end{figure}

\begin{figure}[thb]
\includegraphics[trim = 5mm 0mm 0mm 0mm,scale=0.61]{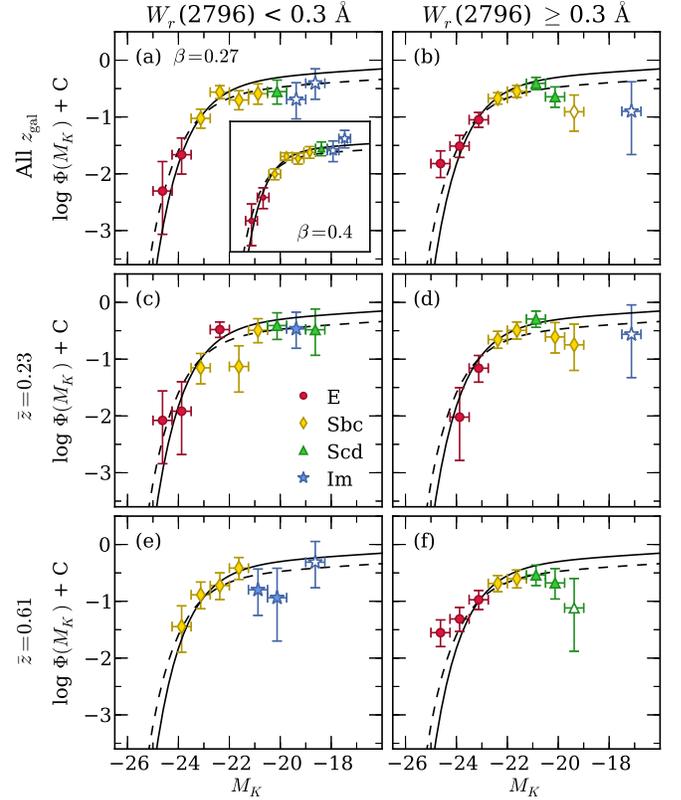}
\caption[]{$K$-band luminosity functions for $W_r(2796) < 0.3$ ($a$,
  $c$, and $e$) and $W_r(2796) \geq 0.3$ ($b$, $d$, and
  $f\,$). Galaxies at all redshifts are included in panels $a$ and
  $b$, while galaxies with $z < \langle z \rangle$ are in panels $c$
  and $d$ (mean $z=0.22$), and galaxies with $z \geq \langle z
  \rangle$ are in panels $e$ and $f$ (mean $z=0.61$), where $\langle z
  \rangle = 0.359$. The solid and dashed curves are $z = 0.3, 1.1$
  Schechter functions, respectively, from \cite{cirasuolo}. Point
  types and colors are the same as described in
  Figure~\ref{fig:Blumfunc}. The inset of panel ($a$) illustrates how
  the observed luminosity function changes with the gas cross section
  correction $\langle R_{\rm gas} \rangle ^{-2} \propto \langle
  L/L_K^{\ast} \rangle^{-2\beta}$.}
\label{fig:Klumfunc}
\end{figure}

For all {\magiicat} galaxies with ``weak'' absorption or
non-detections [$W_r(2796) < 0.3$~{\AA}]\footnote{By examining gas
  defined by $W_r(2796) < 0.3$~{\AA}, whether {\MgII} absorption is
  detected or not detected to the sensitivities afforded by the data,
  we are probing a well-defined gas regime, i.e., gas that does not
  give rise to {\MgII} absorption with $W_r(2796) \geq 0.3$~{\AA}. The
  only assumption is that there is gas probed by the quasar line of
  sight within the region we would consider to be the CGM.}
(Figures~\ref{fig:Blumfunc}$a$ and \ref{fig:Klumfunc}$a$), the
luminosity functions are more or less consistent with those of
\citet{faber} and \citet{cirasuolo}, though there is a trend for a
flattening of the faint-end slopes, especially in $\Phi(M_B)$. The
faint-end slopes of the ``strong'' absorbing galaxies [$W_r(2796) \geq
  0.3$~{\AA}] (Figures~\ref{fig:Blumfunc}$b$ and
\ref{fig:Klumfunc}$b$), are less certain, but suggestive of a
flattening of the faint-end slope relative to the \citet{faber} and
\citet{cirasuolo} luminosity functions.

For the subsamples over the full redshift interval, we find no
statistical differences between the $\Phi(M_B)$ for weak and strong
absorbers (panel $a$ vs.\ $b$ of Figure~\ref{fig:Blumfunc}) as deduced
from a KS test on the unbinned luminosities. The same result applies
for $\Phi(M_K)$ for weak and strong absorbers (panel $a$ vs.\ $b$ of
Figure~\ref{fig:Klumfunc}). Similarly, there are no statistical
differences between the $\Phi(M_B)$ of weak and strong absorbing
galaxies in the low redshift subsample (panel $c$ vs.\ $d$ of
Figure~\ref{fig:Blumfunc}) nor in the $\Phi(M_B)$ in the high redshift
subsample (panel $e$ vs.\ $f$ of Figure~\ref{fig:Blumfunc}). The same
result holds for the $\Phi(M_K)$ of weak and strong absorbing galaxies
for both the low and high redshift subsamples.

For each band, we examined for redshift evolution for the subsample
with weak absorption or non-detections (panels $c$ vs.\ $e$ of
Figures~\ref{fig:Blumfunc} and \ref{fig:Klumfunc}) and for the
subsample with strong absorption (panels $d$ vs.\ $f$ of
Figures~\ref{fig:Blumfunc} and \ref{fig:Klumfunc}). For $\Phi(M_B)$ of
the weak absorbing galaxies, we find redshift evolution at a
$3.4~\sigma$ significance level in the sense that the luminosity
function at low redshift is shifted $\sim 0.5$ magnitudes dimmer
relative to high redshift for galaxies with $M_B<-18$. We find only a
suggestive trend for redshift evolution of $\Phi(M_B)$ of the strong
absorbing galaxies ($2.4~\sigma$). There is also a possible trend for
redshift evolution in the $K$-band of the strong absorbing galaxies
($2.4~\sigma$), but no evidence for the weak absorbing galaxies.

For the $B$-band luminosities, we note that the observed trends from
lower to higher redshift (i.e., flattening of the faint-end slope and
relative overabundance at higher luminosity) are reminiscent of the
Malmquist bias that plagues magnitude-limited surveys. Given the
heterogeneous selection methods used by the various works from which
we constructed {\magiicat}, it is difficult to quantify the degree to
which this may be an issue.

\subsubsection{Color Sequence Along $\Phi(M_B)$ and $\Phi(M_K)$}

The {\magiicat} galaxies (full catalog) exhibit a significant
correlation between $B-K$ and $M_K$ ($7.8~\sigma$), but only a weak
trend between $B-K$ and $M_B$ ($2.0~\sigma$). For the subsample of
strong absorbing galaxies (Figure~\ref{fig:Klumfunc}$b$) the
correlation between $B-K$ and $M_K$ is significant to $6.6~\sigma$ and
for the weak absorbing galaxy subsample (Figure~\ref{fig:Klumfunc}$a$)
the correlation is $3.6~\sigma$. The weak trend between $B-K$ and
$M_B$ is dominated by strong absorbing galaxies ($2.0~\sigma$,
Figure~\ref{fig:Blumfunc}$f$).

The weakest trends (less than $3~\sigma$) between $B-K$ and $M_K$ are
found for the strong absorbing galaxies at low redshift
(Figure~\ref{fig:Klumfunc}$d$) and the weak absorbing galaxies at high
redshift (Figure~\ref{fig:Klumfunc}$e$), even though the mean colors
suggest a strong trend. $B-K$ and $M_K$ are correlated at the
$3.2~\sigma$ level for weak absorbing galaxies at low redshift
(Figure~\ref{fig:Klumfunc}$c$) and $6.4~\sigma$ for strong absorbing
galaxies at high redshift (Figure~\ref{fig:Klumfunc}$f\,$). There is
no correlation between $B-K$ color and $M_B$ for any of the subsamples
presented in Figures~\ref{fig:Blumfunc} except for
Figure~\ref{fig:Blumfunc}$f$, where there is a $3.0~\sigma$
correlation.

The trends and strong correlations between $B-K$ and $M_K$ are
reflected in the color sequence of the data points for $\Phi(M_K)$ in
Figure~\ref{fig:Klumfunc}; the higher the infrared luminosity, the
redder the mean galaxy color. Interestingly, for both the $B$-band and
the $K$-band, KS tests reveal no significant differences in the
distribution of $B-K$ between the weak and strong absorbing galaxies
at low redshift, at high redshift, nor for all redshifts.

Though the overall color distribution may not differ between the
galaxy subsamples, the average SED of the bright end of the $K$-band
luminosity function corresponds to early-type E galaxies and the faint
end corresponds to late-type galaxies (Sbc, Scd, and Im). On the
contrary, there is no clear color differential along the luminosity
sequence of $\Phi(M_B)$. The average color in virtually all $B$-band
luminosity bins is that of an Sbc or Scd SED.

\subsubsection{The $W_r(2796) \geq 0.3$~{\AA} High-Redshift Subsample}

The strong absorbing [$W_r(2796) \geq 0.3$~{\AA}] high redshift
subsamples (Figures~\ref{fig:Blumfunc}$f$ and \ref{fig:Klumfunc}$f\,$)
are most appropriately compared to the SDP94 luminosity functions,
which correspond to $\left< z \right> = 0.65$.

SDP94 find that $\Phi(M_B)$ for {\MgII} absorption selected galaxies
turns down (or ``rolls off'') relative to the faint-end slope of
$\Phi(M_B)$ for field galaxies starting roughly at 1.5 magnitudes
below $M^{\ast}_B$, whereas $\Phi(M_K)$ for {\MgII} absorption
selected galaxies is consistent with $\Phi(M_K)$ of field galaxies to
roughly 3.1 magnitudes below $M^{\ast}_K$ (i.e., $L_K/L^{\ast}_K
\simeq 0.05$). They infer that the absorption selected galaxies
``missing'' from the faint end of $\Phi(M_B)$ are Sd types and later.

SDP94 also report a strong correlation between $B-K$ and $M_K$ in that
fainter galaxies in the $K$-band are bluer, and also find no such
correlation between $B-K$ and $M_B$. Incorporating the correlation
between $B-K$ and $M_K$, they argue that faint blue galaxies
\citep[e.g.,][]{ellis97} do not typically exhibit {\MgII} absorption
with $W_r(2796) \geq 0.3$~{\AA} and that this explains the difference
in the behavior of the faint-end slopes of $\Phi(M_B)$ and
$\Phi(M_K)$. They further conclude that the gas cross section is more
likely to be governed by galaxy mass ($\propto L_K$) than by star
formation. We note that this latter statement is inconsistent with our
finding that the covering fraction of {\MgII} absorption is invariant
with galaxy halo mass for multiple $W_r(2796)$ thresholds
\citep{churchill-masses}.

For {\magiicat} galaxies, the measured $\Phi(M_B)$ for strong
absorbing galaxies at high redshift (Figure \ref{fig:Blumfunc}$f\,$)
is suggestive of a faint-end roll off around 1.5 magnitudes below
$M^{\ast}_B$, consistent with the SDP94 result. Similarly, we find a
possible faint-end roll off in $\Phi(M_K)$ for the strong absorbing
galaxies at high redshift (Figure~\ref{fig:Klumfunc}$f\,$) starting at
roughly 2 magnitudes below $M^{\ast}_K$. This latter result is
contrary to the findings of SDP94. Note that, for this subsample of
the {\magiicat} galaxies, the data extend roughly an additional
magnitude below $M^{\ast}_K$ as compared to the SDP94 sample.

\subsubsection{Interpreting $\Phi(M_B)$ and $\Phi(M_K)$}

Interpreting and comparing the functional forms of $\Phi(M_B)$ and
$\Phi(M_K)$ is rendered difficult due to the heterogeneous selection
methods of the galaxies in {\magiicat}. The majority of galaxies are
``absorption selected'', whereas some galaxies are magnitude-limited
or volume-limited selected. We have applied ``gas cross section''
corrections to all galaxies in the sample, which is a correct
procedure for absorption selected galaxies.

In the cases where galaxies are searched for and identified based upon
prior knowledge of a {\MgII} absorption redshift, there can be
ambiguity as to whether the galaxy is the only galaxy connected to the
{\MgII} absorption. It is always possible that an additional galaxy is
in close projection with the background quasar; given the clustering
properties of galaxies, this would be more rare for two galaxies at
the bright end of the luminosity function, but it could be more
probable as fainter galaxies are considered. These can be found only
with careful point spread function subtraction of the quasar, which
has been performed for all of the quasar fields surveyed from the
ground by GB97 and for the majority of the fields originally surveyed
by SDP94 (in both ground-based and {\it HST\/} images, where the
latter was performed by A.\ Shapley, private communication,
unpublished). However, such analysis has not been performed for the
remainder of the galaxies in {\magiicat}. Furthermore, the absorption
selection approach usually entails an incomplete survey of the
galaxies in the quasar field. If an additional galaxy (or galaxies)
might be discovered in a given quasar field to have a redshift
consistent with the {\MgII} absorption, the galaxies would reclassify
as a ``group'' and would not be included in the present work.

In the cases where galaxies are identified in apparent magnitude- or
volume-limited surveys, the point spread function issue is just as
relevant. Furthermore, apparent magnitude-limited surveys would suffer
from faint-end incompleteness and/or Malmquist bias.

Though the shapes of the luminosity functions presented in
Figures~\ref{fig:Blumfunc} and \ref{fig:Klumfunc} are suggestive of a
relative paucity of sub-$L^{\ast}$ galaxies as compared to field
galaxies, or perhaps even a roll over in the faint-end slopes, the
above considerations make it difficult to assess whether selection
effects are at play. Though the gas cross section corrections we
applied act to reduce the value of $\Phi(M)$ for $L>L^{\ast}$
galaxies, the corrections {\it increase\/} the value of $\Phi(M)$ for
sub-$L^{\ast}$ galaxies; the correction factor increases with
decreasing luminosity. Thus, discrepancies between the observed
luminosity functions and a Schechter function could result from our
not applying the proper cross section correction.

As an exercise, we could attempt to recover the Schechter function
(under the assumption that the luminosity function of {\it all}
galaxies follows this functional form) by varying the $\beta$ in the
gas cross section correction $\langle R_{\rm gas} \rangle ^{-2} \propto
\langle L/L^{\ast} \rangle^{-2\beta}$. In essence, we then learn
something about the gas cross sections of galaxies as a function of
luminosity. In the insets of Figures~\ref{fig:Blumfunc}$a$ and
\ref{fig:Klumfunc}$a$, we provide examples of a $\simeq 0.1$ increase
in $\beta$ for the $W_r(2796) < 0.3$~{\AA} subsamples over all
redshifts. This illustrates that if we have a steeper luminosity
dependence on the gas cross section, especially at the faint end, the
data can be better matched to a Schechter function. That is, matching
the observed luminosity function of absorption selected galaxies to
the Schechter function can, in principle, be used to constrain the
luminosity dependence of the gas cross section. Most importantly,
with an increased sample size (larger than {\magiicat}), we might be
able to determine that the gas cross section does not follow a
constant power law, that the slope $\beta$ is also luminosity
dependent. For example, it is possible to have all subsamples conform
to a Schechter function if we parameterize $\beta$ to have luminosity
dependence such that the gas cross section of low luminosity galaxies
declines more rapidly with decreasing luminosity than it does for high
luminosity galaxies.

Even using the presented luminosity functions, we can still infer
that, in general, galaxy $B-K$ color is independent of galaxy $B$-band
luminosity. Regardless of the $B$-band luminosity, the {\it average\/}
color is consistent with that of an Sbc/Scd galaxy. However, there is
a $B-K$ color sequence in that the greater the infrared luminosity,
the redder the $B-K$ color. This is a highly significant result. If
$M_K$ serves as a very crude proxy for {\it stellar\/} mass, the
luminosity functions suggest that galaxies with lower stellar masses
with detectable {\MgII} absorbing gas comprise bluer (younger) stellar
populations.

This might suggest that a detectable {\MgII} absorbing CGM may not be
present in low stellar mass red galaxies; only the lower stellar mass
galaxies with bluer (younger) stellar populations give rise to
detectable {\MgII} absorption. Since the roll over at the faint-end is
more pronounced for galaxies with $W_r(2796)\geq 0.3$~{\AA} absorbing
gas, we might infer that weaker {\MgII} absorption is preferentially
found in the lower stellar mass galaxies.

\section{Summary and Conclusions}
\label{sec:conclusions}

We compiled, from our own work and the literature, the {\MgII}
Absorbing-Galaxy Catalog, {\magiicat}, consisting of galaxies with
intermediate redshifts to study the galaxy-circumgalactic medium
interaction as probed by {\MgIIdblt} absorption. The catalog presented
here contains 182 isolated galaxies with spectroscopic redshifts
$0.07\leq z\leq 1.1$, impact parameters $D<200$~kpc from a background
quasar, and known {\MgII} absorption or a $3~\sigma$ upper limit on
absorption less than or equal to $0.3$~{\AA}. A summary of the
minimum, maximum, mean, and median values for absorption and galaxy
properties in {\magiicat} is presented in Table~\ref{tab:catprops}.

All values that depend on cosmological parameters have been
recalculated, including quasar-galaxy impact parameters, absolute
magnitudes, and luminosities. This standardizes the galaxy properties
and allows for a comparison of galaxies whose properties were placed
on different cosmologies over the last $\sim 20$ years. Absolute
magnitudes and luminosities were calculated in the $B$-band for all
galaxies and in the $K$-band for all but 18. We find that the average
rest-frame $B-K$ color of {\magiicat} galaxies is consistent with an
Sbc galaxy, though the most common galaxy color is slightly redder
than an Scd galaxy. The average color agrees with SDP94 and
\citet{zibetti07}.

We present the $B$- and $K$-band luminosity functions, $\Phi(M)$, for
subsamples split by $W_r(2796)<0.3$~{\AA} (``weak'') and $W_r(2796)
\geq 0.3$~{\AA} (``strong''), and by low redshift and high redshift,
cut by $z=0.359$. We find that the luminosity functions in both bands
for galaxies with weak absorption or non-detections are more-or-less
consistent with \citet{faber} and \citet{cirasuolo}, while the strong
absorbing galaxies may be flatter in the faint-end slope. Comparing
the strong, high redshift subsample to SDP94, we find the suggestive
faint-end roll off of the luminosity function in the $B$-band
consistent, while the possible roll off in the $K$-band is
contrary. No statistical difference between weak and strong subsamples
for all, low redshift, and high redshift galaxies is present in either
the $B$- or the $K$-band. The $B$-band may show redshift evolution in
both the weak and strong subsamples, but in the $K$-band, evolution
may only be present in the strong subsample. These above statements
depend upon the gas cross section correction factor that we have
applied to the data. We discussed how the luminosity functions can,
in principle, be used to constrain the dependence of the gas cross
section on galaxy luminosity.

We find a correlation between $B-K$ and $M_K$ for the full sample
($7.8~\sigma$), but only a weak correlation between $B-K$ and $M_B$
($2.0~\sigma$), consistent with the findings of SDP94. Splitting
{\magiicat} into weak absorbers, strong absorbers, low redshift, and
high redshift subsamples, we find the correlations in both bands are
dominated by the high redshift, strong absorbing galaxies. As $M_K$
becomes brighter, the mean galaxy color becomes redder. On the other
hand, the mean color of most magnitude bins in the $B$-band luminosity
functions is consistent with an Sbc/Scd SED.

The behavior of the luminosity functions suggest that only the lower
stellar mass galaxies with bluer (younger) stellar populations give
rise to detectable {\MgII} absorption. Comparing the faint-end roll
over between galaxies with $W_r(2796)\geq 0.3$~{\AA} and
$W_r(2796)<0.3$~{\AA} absorbing gas, it would seem that in lower
stellar mass galaxies, weaker {\MgII} absorption would be observed
more commonly.

Further analysis has already been conducted with all or a portion of
the galaxies in {\magiicat}. \citet{kcn12} studied the effect of
galaxy orientation on {\MgII} absorption. They found that {\MgII} gas
is preferentially found along the galaxy major and minor axes, where
the covering fraction of {\MgII} absorption as a function of
orientation is enhanced by as much as $20\%-30\%$ along the major and
minor axes. This bimodality was found to be driven by blue galaxies
and may indicate outflowing gas with an opening angle of $100^{\circ}$
and inflowing gas with an opening angle of $40^{\circ}$.

Using halo abundance matching, \citet{churchill-masses, cwc-masses2}
obtained the galaxy virial masses for all galaxies in {\magiicat} and
studied the $W_r(2796)$-$D$ anti-correlation and {\MgII} covering
fractions, both as a function of galaxy virial mass. They found that
the {\MgII} CGM has projected absorption profile that follows
$(D/R_{\rm vir})^{-2}$, where $R_{\rm vir}$ is the virial radius,
indicating a self-similar behavior with virial mass. They also find
that $W_r(2796)$ increases with virial mass in finite ranges of $D$
but is constant in finite ranges of $D/R_{\rm vir}$, and that covering
fractions are unchanged as a function of galaxy virial mass within a
given $D$ or $D/R_{\rm vir}$. Their results are contrary to the
theoretical prediction that cold-mode accretion is shut down in
high-mass galaxies \citep{birnboim03, dekel06, stewart11}.

\citet{nielsen12} [Paper II of this series] presented an analysis of
how $W_r(2796)$, covering fractions, and the radial extent of the
{\MgII} CGM depend on impact parameter, galaxy redshift, $B$- and
$K$-band luminosities, and $B-K$ color. They found that the
anti-correlation between $W_r(2796)$ and $D$ can be characterized by a
log-linear fit which levels off at low $D$. The scatter on the
$W_r(2796)-D$ plane may be due to galaxy luminosity, where more
luminous galaxies have larger $W_r(2796)$ at a fixed $D$. They also
found that the covering fraction decreases with increasing $D$ and
increasing $W_r(2796)$ threshold. More luminous, bluer, and higher
redshift galaxies have larger covering fractions than less luminous,
redder, and lower redshift galaxies at a given $D$. The
luminosity-scaled radial extent of the {\MgII} CGM is more sensitive
to luminosity in the $B$-band than in the $K$-band. The radial extent
has a steeper luminosity dependence for red galaxies than blue
galaxies, and for low redshift than high redshift galaxies.

In future work we intend to apply multivariate analysis methods to
{\magiicat}, incorporating the galaxy virial mass estimates of the
galaxies from \citet{cwc-masses2} as well as {\MgII} kinematics
and the low- and high-ionization absorption strengths of the CGM. We
also plan to utilize the sample of group galaxies we obtained with the
present work, comparing the group galaxies to the isolated
galaxies. Mining the {\it Hubble Space Telescope\/} archive for {\HI}
and UV low- and high-ionization metal-line transitions will be useful
for developing a more complete understanding of the CGM properties of
{\magiicat} galaxies.

\acknowledgments 

We thank the referee for helpful comments which improved the
manuscript. We also thank Jacqueline Bergeron for providing details on
the observations conducted in \citet{guillemin}, Chuck Steidel for
communicating detailed aspects relating to the work by \citet{sdp94},
including access to unpublished materials, and David Law for
additional magnitudes. We extend our gratitude to all the researchers
whose hard work over the last two decades resulted in the growing
database of published galaxies in quasar fields. Without their
efforts, this work would not have been possible. This research was
primarily supported through grant HST-AR-12646 provided by NASA via
the Space Telescope Science Institute, which is operated by the
Association of Universities for Research in Astronomy (AURA) under
NASA contract NAS 5-26555. This work was also supported by the
Research Enhancement Program provided by NASA's New Mexico Space Grant
Consortium (NMSGC). N.M.N.~was also partially supported through a
NMSGC Graduate Fellowship and a Graduate Research Enhancement Grant
(GREG) sponsored by the Office of the Vice President for Research at
New Mexico State University. M.T.M.~thanks the Australian Research
Council for a QEII Research Fellowship (DP0877998) and Discovery
Project grant (DP130100568). This research has made extensive use of
the SAO/NASA Astrophysics Database System (ADS), operated by the
Smithsonian Astrophysical Observatory under contract with NASA, the
NASA/IPAC Extragalactic Database (NED), operated by the Jet Propulsion
Laboratory and California Institute of Technology, under contract with
NASA, and the SIMBAD database, operated at Centre de Donn\'{e}es,
Strasbourg, France. This research has also made use of the Spanish
Virtual Observatory (http://svo.cab.inta-csic.es) supported from the
Spanish MICINN / MINECO through grants AyA2008-02156, AyA2011-24052

\begin{appendix}

\section{$K$-corrections}
\label{app:kcorr}

For a galaxy at redshift $z$ and observed in bandpass $y$, the
$K$-correction between bandpass $y$ and desired bandpass $x$ is:
\begin{equation}
K_{xy}=2.5\log(1+z) +2.5\log \left[ \frac{\int R_x(\lambda)\, \lambda
    \,f_{\lambda}(\lambda) \, \,d\lambda}{\int R_y(\lambda)\, \lambda
    \,f_{\lambda}(\lambda /[1+z])\, \,d\lambda}\right]
+2.5\log\left[\frac{\int R_y(\lambda)\, \lambda
    \,f_{\lambda}^{\,s}(\lambda)\, \,d\lambda}{\int R_x(\lambda)\,
    \lambda \,f_{\lambda}^{\, s}(\lambda)\, \,d\lambda}\right],
\label{eq:kcorr}
\end{equation} 
where $R_y(\lambda)$ is the response curve of the $y$-band,
$R_x(\lambda)$ is the response of the $x$-band, $f_{\lambda}(\lambda)$
is the flux density of the object being observed in the object's rest
frame, $f_{\lambda}(\lambda /[1+z])$ is the flux density of the
redshifted object in the observer frame, and $f_{\lambda}^{\,
  s}(\lambda)$ is the standard Vega or AB spectrum. The first two
terms of the $K$-correction correct for the fact that the observed
object's spectrum is stretched and shifted redwards at larger $z$. The
last term is the color term which corrects for different observed
($y$) and desired ($x$) bandpasses. If these bandpasses are identical,
the color term cancels out.

\begin{figure}[ht]
\centering 
\includegraphics[angle=0,trim = 3mm 0mm 0mm 0mm,scale=0.9]{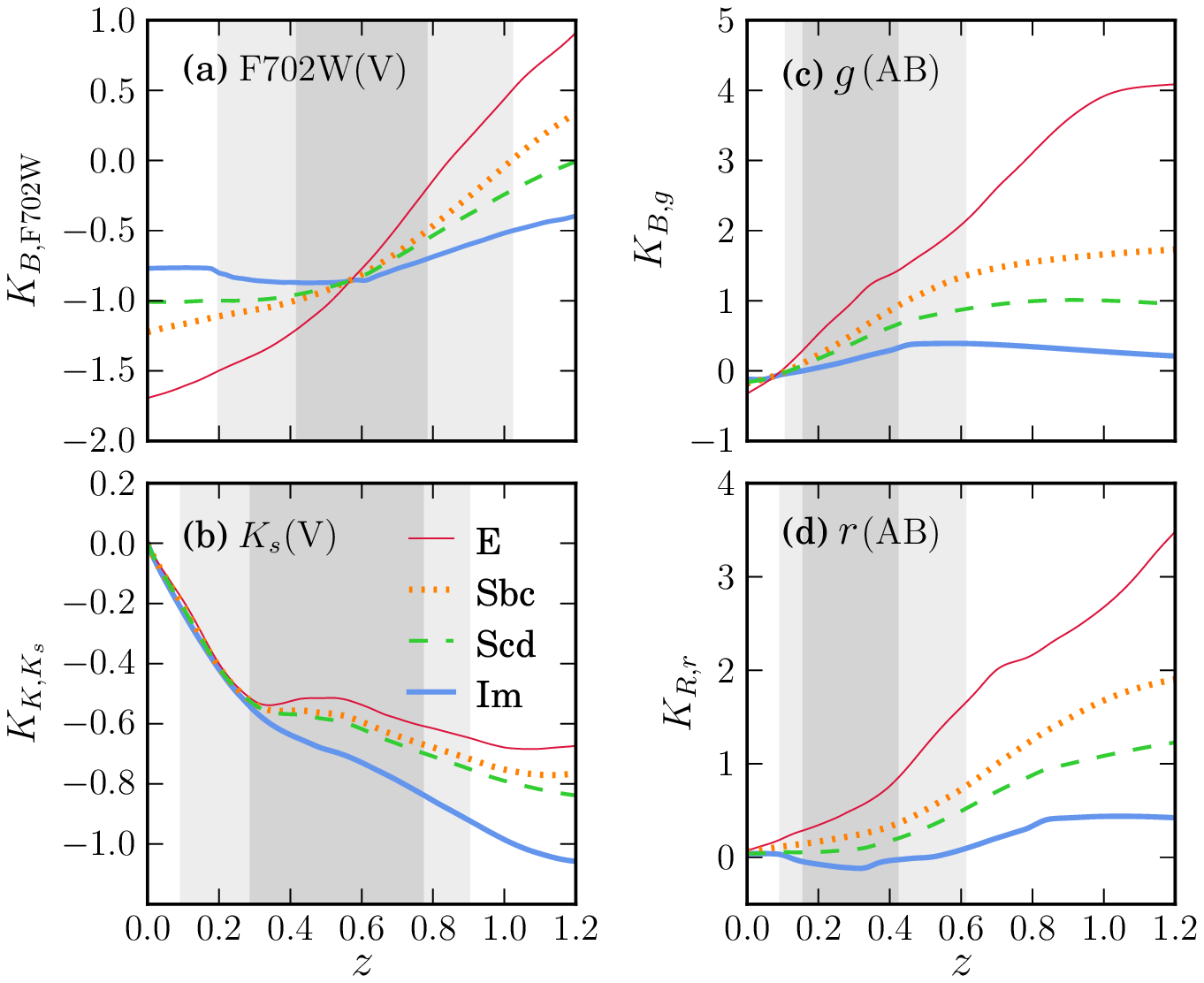}
\caption[]{The $K$-corrections between various bands as a function of
  redshift for spectral energy distributions E, Sbc, Scd, and
  Im. Apparent magnitudes in the F702W(Vega) and $g$(AB) bands were
  $K$-corrected to the $B$-band, $K_s$(Vega) to the $K$-band, and
  $r$(AB) to the $R$-band. The lighter gray shading indicates the full
  redshift region over which the $K$-corrections were applied while
  the darker gray shading indicates where the central $68\%$ of the
  galaxies reside.}
\label{fig:kcorrs}
\end{figure}

We obtained the required filter response curves, $R_x(\lambda)$ and
$R_y(\lambda)$, from the website associated with the facility used to
observe a given galaxy, or from the author of the work. Chuck Steidel
(private communication) kindly provided us both the $R_s$ and the
$K_s$ response curves while Jacqueline Bergeron (private
communication) provided the $R_{\hbox{\tiny EFOSC}}$ response
curve. We also used the Spanish Virtual Observatory Filter Profile
Service\footnote{http://svo2.cab.inta-csic.es/theory/fps/}. We
retrieved the Vega composite flux standard spectrum
``alpha\_lyr\_stis\_005'' from the STScI Calibration Database System,
Calspec\footnote{
  http://www.stsci.edu/hst/observatory/cdbs/calspec.html}. For AB
magnitudes, we calculated a standard AB spectrum, defined as a
hypothetical source with $F_{\nu}^{(\rm AB)}=3.63x10^{-20}$ [erg
  s$^{-1}$ cm$^{-2}$ Hz$^{-1}$].

Figure~\ref{fig:kcorrs} presents the $K$-correction for each SED as a
function of redshift for Vega magnitudes F702W and $K_s$, and AB
magnitudes $g$ and $r$, the four most common bandpasses in
{\magiicat}. Vega magnitudes F702W and $K_s$ were corrected to Vega
magnitudes $B$ and $K$, respectively, while AB magnitudes $g$ and $r$
were corrected to AB magnitudes $B$ and $R$, respectively. The lighter
gray shading indicates the full galaxy redshift range over which the
$K$-corrections were applied for that band. The darker gray shading
indicates where the central $68\%$ of the galaxies reside.

We determined the constants of conversion between Vega and AB
magnitudes by calculating the $B$- and $K$-band absolute magnitudes
for a given SED alternately using the Vega spectrum and the AB
spectrum. Taking the difference of the Vega and AB magnitudes in each
band produced the conversions $B({\rm AB})=B({\rm V})-0.0873$ and
$K({\rm AB})=K({\rm V})+1.8266$. These constants are comparable to the
values $-0.09$ ($B$-band) and $1.85$ ($K$-band) which are presented in
Table~1 of \citet{blanton07}. We applied these constants to the
$K$-corrected apparent magnitudes when necessary.

\section{Selecting Spectral Energy Distributions for $K$-corrections}
\label{app:SEDcolors}

For $f_{\lambda}(\lambda)$ in Equation~\ref{eq:kcorr}, we did not have
direct access to the galaxy spectra for {\magiicat}
galaxies. Therefore we rely on \citet{cww80} spectral energy
distribution (SED) templates which were extended to shorter and longer
wavelengths by \citet{bolzonella00} using synthetic spectra created
with the {\sc gissel98} code \citep{bc93}. These SEDs were distributed
with and used by the {\sc
  hyperz}\footnote{http://webast.ast.obs-mip.fr/hyperz/} photometric
redshift code \citep{bolzonella00}.

To select which $K$-correction to apply, we determined which SED each
galaxy in {\magiicat} most closely resembles. We compared the observed
colors of the galaxy to each SED type at the galaxy's redshift. In
cases where the observed galaxy color was in between the color of two
different SEDs, the closest SED was selected. Where no observed colors were
available due to a lack of a second band, the galaxy was classified as
an Sbc galaxy, the average type for {\MgII} absorbing galaxies
\citep[SDP94;][]{zibetti07}.

We calculated the SED observed colors by first determining the
rest-frame colors of each SED. This was done by calculating the
appropriate apparent magnitudes of the SEDs at $z=0$ and taking the
difference of the magnitudes, making sure to use the correct filter
and AB/Vega combinations. We then calculated the necessary
$K$-corrections at redshifts $0<z<1.2$ with the methods described in
Appendix~\ref{app:kcorr}. To obtain the observed SED colors for the
redshift range of {\magiicat}, we combined these $K$-corrections with
the rest-frame SED colors, e.g., $({\rm F702W}-K_s)_{z>0}=({\rm
  F702W}-K_s)_{z=0}+[K_{Ks}(z)-K_{F702W}(z)]$, where the terms in the
square brackets are the $K$-corrections for the $K_s$ and F702W bands,
respectively, for $z>0$.

\begin{figure}[th]
\centering
\includegraphics[angle=0,trim = 3mm 0mm 0mm 0mm,scale=0.8]{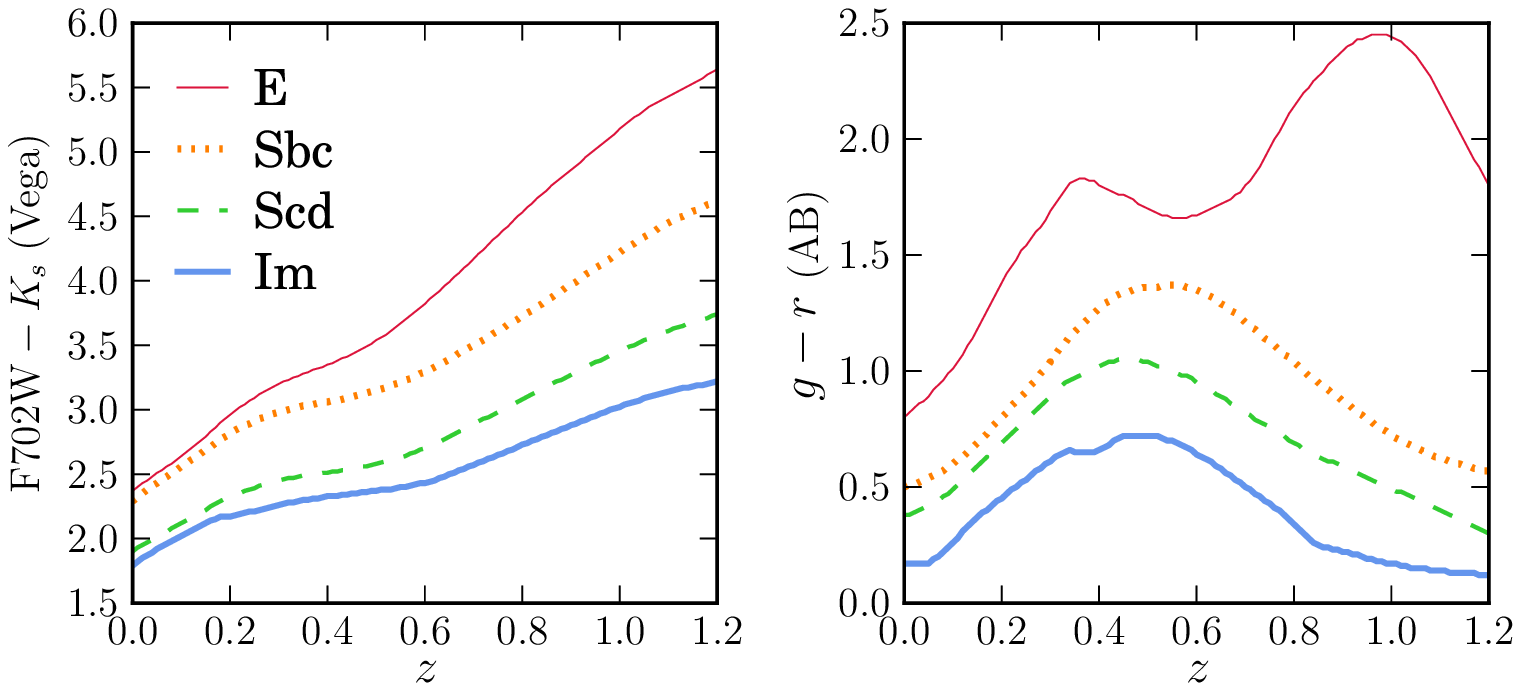}
\caption[]{Observed colors, ${\rm F702W}-K_s$ (Vega) and $g-r$ (AB),
  for the galaxy SEDs as a function of redshift. These two colors are
  the most common observed colors in {\magiicat}. The rest of the
  colors follow the same trend as ${\rm F702W}-K_s$.}
\label{fig:SEDcolors}
\end{figure}

The two most common observed colors in the sample are ${\rm
  F702W}-K_s$ (Vega), and $g-r$ (AB). These colors are presented in
Figure~\ref{fig:SEDcolors} for all SEDs. The rest of the observed
colors follow the trend of ${\rm F702W}-K_s$ due to a combination of a
red band with an infrared band. Our $g-r$ colors are consistent with
\citet{hewett06}, who present $g-r$ for the extended Coleman SEDs.

\section{$B-R$ to $B-K$ Rest-Frame Color Conversion}
\label{app:BKBR}

For SDSS galaxies which did not have a $K$-band magnitude available,
we determined $M_K$ indirectly by using a conversion between
rest-frame colors $B-R$ and $B-K$. We calculated the rest-frame $B-K$
and $B-R$ colors for each SED using the methods applied in
Appendix~\ref{app:SEDcolors}. The points in Figure~\ref{fig:BKBR}
present the rest-frame colors for each SED. The SED colors appear to
follow a linear trend so a linear least-squares fit was performed. The
fit has the form $(B-K)=1.86(B-R)+0.02$ and is presented as the solid
line in Figure~\ref{fig:BKBR}. We therefore obtained absolute $K$-band
magnitudes, $M_K$, for each SDSS galaxy by applying the equation
$M_K=M_B-(B-K)=M_B-1.86(B-R)-0.02$.

\begin{figure}[th]
\centering
\includegraphics[angle=0,trim = 3mm 0mm 0mm 0mm,scale=0.8]{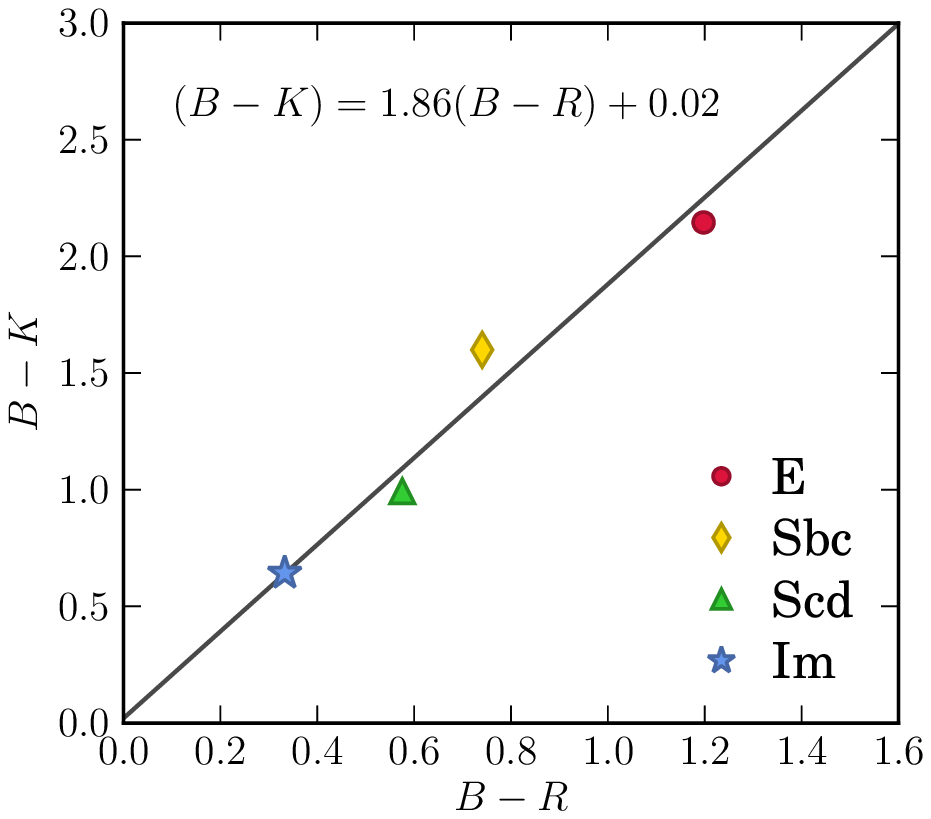}
\caption[]{The relationship between rest-frame colors $B-K$ and $B-R$
  for the SEDs. Point types and colors indicate the SED type. The data
  appear to follow the linear relation listed on the figure.}
\label{fig:BKBR}
\end{figure}

\end{appendix}


\end{document}